\documentclass[fleqn,usenatbib]{mnras}

\usepackage{newtxtext,newtxmath}

\usepackage[normalem]{ulem}
\usepackage{soul}
\usepackage{graphicx}
\usepackage{rotating}
\usepackage[T1]{fontenc}
\usepackage{booktabs}
\usepackage{lscape}
\DeclareRobustCommand{\VAN}[3]{#2}
\let\VANthebibliography\thebibliography
\def\thebibliography{\DeclareRobustCommand{\VAN}[3]{##3}\VANthebibliography}

\usepackage{graphicx}	
\usepackage{amsmath}	
\usepackage{amssymb}	

\title[Deuterium fractionation in L1688]{Deuterium fractionation in cold dense cores in the low-mass star forming region L1688\footnote{This work is based on observations carried out under projects 009-15, and 031-16 with the IRAM~30~m telescope. IRAM is supported by INSU/CNRS (France), MPG (Germany) and IGN (Spain).}}

\author[I. V. Petrashkevich et al.]{
I. V. Petrashkevich$^{1}$\thanks{E-mail: petrashkevich.igor@gmail.com},
A. F. Punanova$^{2}$,
P. Caselli$^{3}$,
O. Sipil\"a$^{3}$,
J. E. Pineda$^{3}$,
R. K. Friesen$^{4}$,
\newauthor
M. G. Korotaeva$^{1}$,
A. I. Vasyunin$^{1}$
\\
$^{1}$Research Laboratory for Astrochemistry, Ural Federal University, Mira st. 19, 620002 Yekaterinburg, Russia\\
$^{2}$Onsala Space Observatory, Chalmers University of Technology, Observatoriev\"agen 90, R\aa\"o, Onsala, Sweden \\
$^{3}$Max Planck Institute for extraterrestrial Physics, Giessenbachstrasse 1, 85748 Garching, Germany \\
$^{4}$Department of Astronomy \& Astrophysics, University of Toronto, 50 St. George St., Toronto, ON M5S 3H4, Canada
}

\date{Accepted XXX. Received YYY; in original form ZZZ}

\pubyear{2023}

\begin{document}
\label{firstpage}
\pagerange{\pageref{firstpage}--\pageref{lastpage}}
\maketitle

\begin{abstract}
In this work, we study deuterium fractionation in four starless cores in the low-mass star-forming region L1688 in the Ophiuchus molecular cloud. We study how the deuterium fraction ($R_D$) changes with environment, compare deuteration of ions and neutrals, core centre and its envelope, and attempt to reproduce the observed results with a gas-grain chemical model.
We chose high and low gas density tracers to study both core centre and the envelope. With the IRAM 30~m antenna, we mapped N$_2$H$^+$(1--0), N$_2$D$^+$(1--0), H$^{13}$CO$^+$ (1--0) and (2--1), DCO$^+$(2--1), and $p$-NH$_2$D(1$_{11}$--1$_{01}$) towards the chosen cores. The missing $p$-NH$_3$ and N$_2$H$^+$(1--0) data were taken from the literature. To measure the molecular hydrogen column density, dust and gas temperature within the cores, we used the Herschel/SPIRE dust continuum emission data, the GAS survey data (ammonia), and the COMPLETE survey data to estimate the upper limit on CO depletion. 
We present the deuterium fraction maps for three species towards four starless cores. Deuterium fraction of the core envelopes traced by DCO$^+$/H$^{13}$CO$^+$ is one order of magnitude lower ($\sim$0.08) than that of the core central parts traced by the nitrogen-bearing species ($\sim$0.5). Deuterium fraction increases with the gas density as indicated by high deuterium fraction of high gas density tracers and low deuterium fraction of lower gas density tracers and by the decrease of $R_D$ with core radii, consistent with the predictions of the chemical model. 
Our model results show a good agreement with observations for $R_D$(N$_2$D$^+$/N$_2$H$^+$) and R$_D$(DCO$^+$/HCO$^+$) and underestimate the  $R_D$(NH$_2$D/NH$_3$).
\end{abstract}




\section{Introduction}

Pre-stellar cores are cold, dense, quiescent \citep[$T\simeq10$~K, $n$(H$_2$)=10$^4$--10$^7$~cm$^{-3}$, e.g.,][]{Ward-Thompson1999,Pineda2023} structural elements of molecular clouds. Under these physical conditions, CO, the most abundant molecule after H$_2$ in cold gas, freezes onto dust grains. With low abundance of CO in the gas phase, deuteron-proton exchange reaction, that starts deuteration, can proceed efficiently \citep{Dalgarno1984}, which leads to an increase in deuterium fraction \citep[e.g.,][]{Crapsi2005}. Deuterium fraction becomes an important instrument to study the pre-stellar phase; it is measured as a column density ratio of deuterated and hydrogenated isotopologues \citep[e.g.][]{Bacmann2003,Crapsi2005}.
Study of deuterium fractionation is important to understand the chemical processes in pre-stellar cores, the first stages of star formation, and to put constraints on chemical models \citep[e.g.,][]{Caselli2012,Ceccarelli2014}.

In molecular clouds, at high gas densities and low temperatures, various species are adsorbed onto the dust grain surfaces depending on their desorption energies and formation routes. Carbon-bearing species form early in the molecular cloud and are known to start depleting catastrophically at gas densities above 10$^4$~cm$^{-3}$ \citep{Caselli1999}. Some nitrogen-bearing species, such as NH$_3$ and N$_2$H$^+$, need molecular nitrogen to be formed, that is they need longer chemical evolution of the cloud \citep{Hily-Blant2010}. \citet{Hily-Blant2010} showed that at densities of $10^4$~cm$^{-3}$, the typical time scale to convert N to N$_2$ is of the order of 10$^6$~yr, while 
the chemical age of a typical pre-stellar core is $\sim10^5$~yr \citep[e.g.,][]{Tafalla2004,Jimenez-Serra2021,Punanova2022}. This implies that there is a large gas-phase N$_2$ reservoir to form N-bearing species in pre-stellar cores, while N$_2$ is actively formed. Thus such C-bearing species as HCO$^+$ and their isotopologues mostly trace a less dense gas of core shell, where CO is abundant, while such N-bearing species as N$_2$H$^+$ and NH$_3$, the products of N$_2$, mostly trace dense core gas. In centrally concentrated prestellar cores, N-bearing species are also expected to freeze out onto dust grains deep inside (within the central few thousand a.u.), leaving only species lighter than He in the gas phase \citep{Caselli2022}. Measuring the deuterium fraction ($R_D$) in different tracers gives us an independent way to study deuterium fractionation under different physical conditions, in addition to the direct study of the correlations between the deuterium fraction and the physical conditions (gas density, gas and dust temperature) or linear scale (distance to the dust peak) within the cores.

Deuterium fraction in cold dense cores is typically measured as a column density ratio of deuterated and hydrogentaed isotopologues in many observational studies, most of them being single pointings \citep[one core, one measurement using a single-dish telescope, e.g.,][]{Bacmann2003,Crapsi2005,Friesen2013,Punanova2016,Harju2017,Ambrose2021,Bovino2021} and some presented maps \citep{Caselli2002,Parise2011,Friesen2010_3,Chacon-Tanarro+2019}. The previous studies show that deuterium fraction in dense cores increases up to 0.7 \citep{Pagani2007} while the elemental abundance of deuterium in Solar neighbourhood is $1.5\times10^{-5}$ \citep{Linsky2006}. The highest deuteration is observed towards starless cores in low-mass star-forming regions \citep[$R_D$(N$_2$H$^+$)$\le$0.5; e.g.,][]{Crapsi2005,Punanova2016}. The species that are abundant in the gas of core envelopes show smaller deuterium fractions \citep[e.g., 0.005--0.080 in HCO$^+$, H$_2$CO, CH$_3$OH;][]{Caselli2002,Tafalla2006,Bergman2011,Chacon-Tanarro+2019,Redaelli2019,Ambrose2021} then those abundant in the core centres \citep[0.02--0.70 in N$_2$H$^+$, NH$_3$,][]{Crapsi2005,Pagani2007,Punanova2016,Harju2017,Redaelli2019}.

 The target of our study, L1688, is a site of low-mass star formation, a part of the $\rho$-Ophiuchus molecular cloud, where active star formation takes place \citep{Loren1990}. L1688 is located at a distance of 119~pc \citep{Zucker2019}, which makes it a good target for an observational study. This region contains tens of starless and protostellar cores \citep[e.g.,][]{DiFrancesco2008}. \citet{Frisen2017} and \citet{Choudhury2021} evaluated the gas temperature in all L1688 through ammonia observations and found it to be in a range of 10~K (towards cold dense cores) to 30~K (towards the clumps with protostars). \citet{Ladjelate2020} showed the dust temperature (12--24~K) of the entire L1688 region based on the Herschel/SPIRE data. L1688 consists of several clumps, A, B, C, D, E, F, H, I that have different size, mass, number of dense cores and embedded protostars, temperature, level of turbulence \citep{Motte1998,Andre2007,DiFrancesco2008,Frisen2017}.
 
 Deuterium fractionation in L1688 was studied in several works. \citet{Parise2011} {and \citet{Bovino2021}} studied o-H$_2$D$^+$/p-D$_2$H$^+$ towards the pre-stellar core Oph-H-MM1 and other cores in Ophiuchus. \citet{Friesen2010_3} presented the deuterium fraction map in N$_2$D$^+$/N$_2$H$^+$ for Oph-B2, a star-forming clump with cold dense cores and embedded protostars. \citet{Harju2017} presented a detailed study of deuteration in N$_2$H$^+$ and NH$_3$ towards Oph-H-MM1 based on single-pointing observations of all their possible isotopologues including doubly and triply deuterated ammonia. \citet{Punanova2016} presented single-pointing observations of N$_2$H$^+$, N$_2$D$^+$, and C$^{17}$O towards 40 dense cores. They revealed high level of deuteration ($R_D$=0.02--0.43) and low level of CO depletion in all sub-regions of L1688. Some of the dense cores showed very high deuteration with very low CO depletion which contradicts the current understanding of deuterium chemistry in cold cores \citep[$R_D$=0.30--0.43 with CO depletion $\leq 4.4$;][]{Punanova2016}. 
 
We present for the first time a spatial distribution of deuterium fraction in four dense cores in three different pairs of species, N$_2$H$^+$ and N$_2$D$^+$, NH$_3$ and NH$_2$D, H$^{13}$CO$^+$ and DCO$^+$, and search for correlations between deuterium fractions and physical parameters of the gas. We consider the N-bearing species as tracers of dense gas and the C-bearing species as tracers of less dense core envelopes. For the mapping, we chose four of the highly deuterated cold dense cores that show low CO depletion: Oph-C-N, Oph-E-MM2, Oph-F, Oph-H-MM1, located in four sub-regions of L1688, three of them located far from protostars, to avoid the protostellar feedback. We also do a single-pointing probe towards the cores Oph-H-MM2 and Oph-I-MM2, newly identified via ammonia observations \citep{Frisen2017}. The mapped cold dense cores are stationary, bound by pressure \citep[the virial parameters of the cores were estimated in][]{Pattle2015,Kerr2019,Singh2021}.). We model the chemical composition for the studied cold dense cores using the gas-grain chemical model pyRate3 presented in \citet{Sipila2015_model,Sipila2015_spin_state} and compare two different approaches to reproduce the physical profiles of the cores and two initial abundances of nitrogen.

Section~\ref{sec:observations} describes the observations presented in this work and other observations data taken from the literature. Section~\ref{sec:data_reduction} describes the processing of observational data and analysis of the spectral lines. Section~\ref{sec:results} describes the results of the line analysis, the obtained column densities of the species and the deuterium fraction. Section \ref{sec:model} describes the chemical modeling of the studied dense cores and the comparison of the model results with the observations. In Sect.~\ref{sec:discussion}, we discuss our line analysis method, the obtained deuterium fraction values, the difference in deuterium fraction between the tracers, and the performance of the chemical model.  In Section~\ref{sec:conclusions} we present our conclusions.

\section{Observations}\label{sec:observations}
\begin{figure*}
\centering
\includegraphics[scale=0.38]{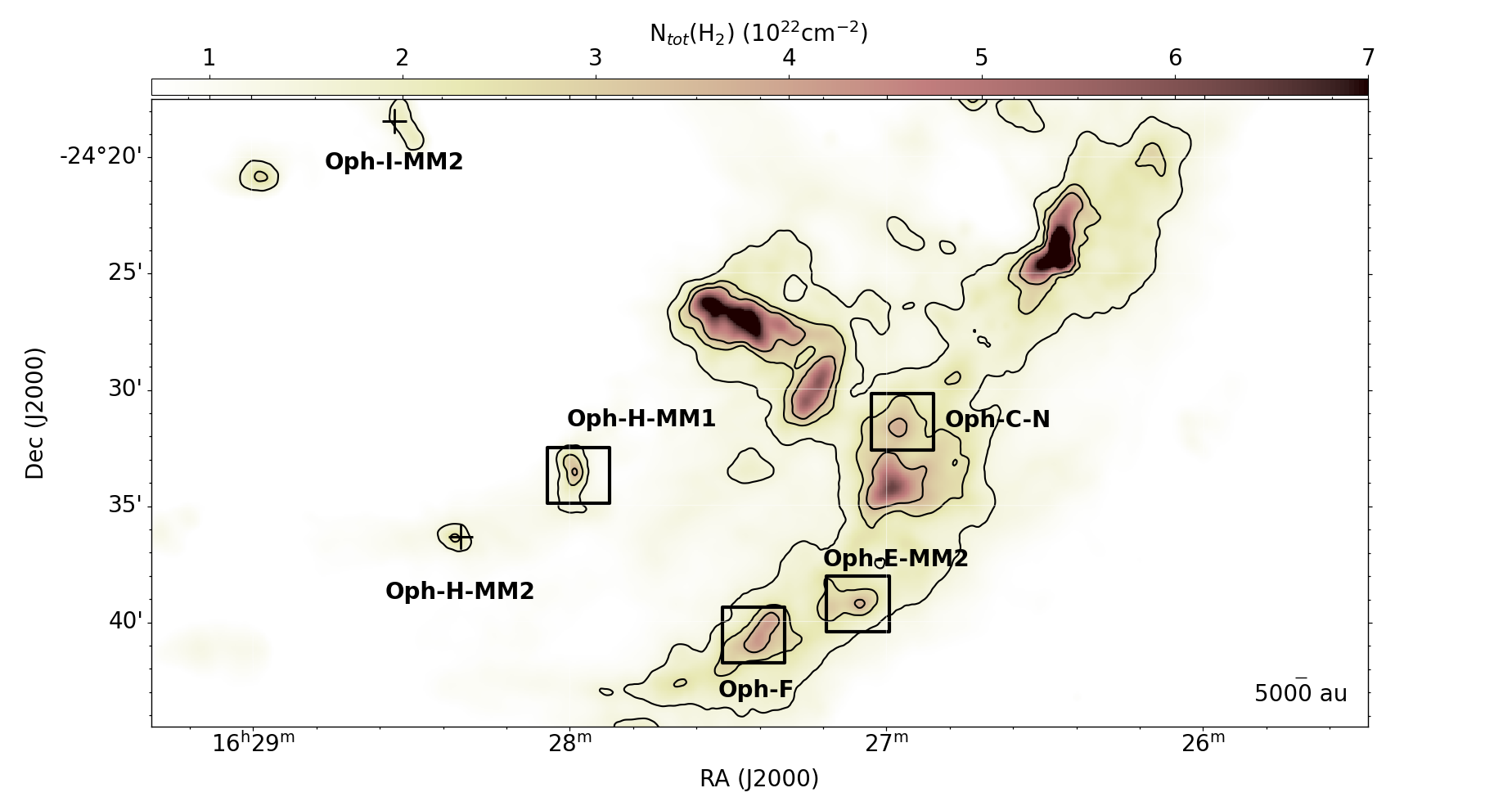}
\caption{The L1688 star-forming region as seen in the molecular hydrogen column density ($N_{\rm tot}$(H$_2$), color scale) from \citet{Ladjelate2020}. The first contour starts at 1.5$\times$10$^{22}$~cm$^{-2}$  with a contour step of 1.0$\times$10$^{22}$~cm$^{-2}$. Black rectangles show mapped cores and black crosses show single-pointing observations. }
\label{pic:L1688}
\end{figure*}

We mapped four cold dense cores: Oph-C-N, Oph-E-MM2, Oph-F, Oph-H-MM1, located in the L1688 star-forming region and performed single-pointing observations towards Oph-H-MM2 and Oph-I-MM2, newly detected by \citet{Frisen2017}. The cores are shown in Fig.~\ref{pic:L1688} and their coordinates are listed in Table~\ref{tab:coord}. The observations were carried out with the IRAM-30~m telescope on June 26 -- July 6 2015 (IRAM project 009-15) and on September 20--23 2016 (IRAM project 031-16). We observed the DCO$^+$(2--1), H$^{13}$CO$^+$ (1--0) or (2--1), p-NH$_2$D(1$_{11}$--$1_{01}$), N$_2$D$^+$ (1--0) or (2--1), and N$_2$H$^+$(1--0) lines. The on-the-fly mapping and single-pointing observations were done simultaneously in 2~mm and 3~mm bands, in position-switching mode, under acceptable weather conditions ($pwv$ = 5--10~mm). The off-positions were observed once in $\leq$10~minutes. The telescope was equipped with the EMIR090 and EMIR150 receivers and the VESPA spectrograph, with a spectral resolution of 20 kHz that corresponds to 0.04--0.08 km~s$^{-1}$ at the observed frequencies. The sensitivity of the observations was 0.05--0.60~K ($T_{\rm mb}$). The exact transition frequencies, spectral resolutions, beam sizes, antenna efficiencies, system temperatures, and weather conditions are listed in Table~\ref{tab:allline}. The telescope pointing was checked once in two hours by observing QSO~1757-240, QSO~1730-130, QSO~1253-055, QSO~1514-241, QSO~1622-297, Saturn, or Venus. The telescope focus was checked once in six hours by observing Saturn, Venus, QSO~1757-240, or QSO~1253-055.

\subsection{The data from the literature}
To measure deuterium fraction, we used the data on the hydrogenated isotopologues where they were available in the literature. The N$_2$H$^+$(1--0) data for Oph-F and Oph-E-MM2, observed with the IRAM~30~m antenna and SIS heterodyne receiver with an autocorrelation spectrometer as backend, were taken from \citet{Andre2007}. These observations have spectral resolution 20--40~kHz and the beam size 26.4$^{\prime\prime}$, rms$\simeq$0.1--0.2~K ($T_{\rm mb}$). The p-NH$_3$ (1,1) and (2,2) observations for Oph-C-N, Oph-E-MM2, and Oph-F observed with the GBT telescope (32$^{\prime\prime}$ beam) and the KFPA spectrograph, were taken from the GAS survey by \citet{Frisen2017}. These observations have spectral resolution of 5.7~kHz and rms of 0.01~K ($T_{\rm mb}$). To search for the correlations between deuterium fraction and physical conditions within the cores, and to construct the core profiles for modelling, we used dust temperature and molecular hydrogen column density maps based on Herschel/SPIRE observations from \citet{Ladjelate2020}, with effective resolution of 18.2$^{\prime\prime}$. The gas temperature was also taken from the GAS survey \citep{Frisen2017}, as mentioned above. The $^{13}$CO(1--0) observations of the COMPLETE survey \citep{Ridge2006} obtained with the FCRAO antenna (46$^{\prime\prime}$~beam and 0.07~km~s$^{-1}$ spectral resolution) were used to estimate the lower limit for the CO column density and the upper limit for the CO depletion.  

\begin{figure*}
\center{\includegraphics[width=1\linewidth]{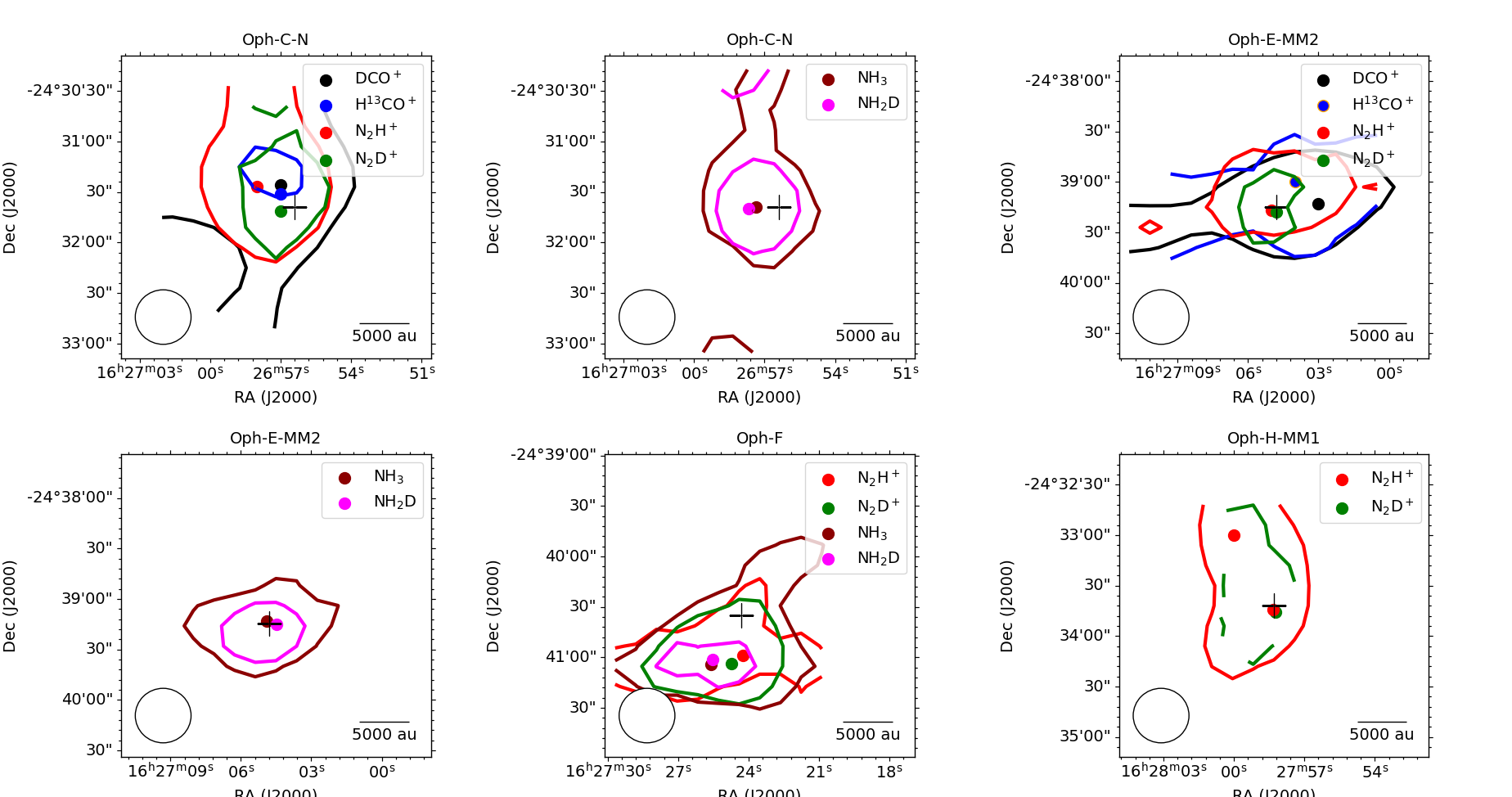}}
\caption{Integrated intensities of the DCO$^+$(2--1), H$^{13}$CO$^+$(1--0) or H$^{13}$CO$^+$(2--1), NH$_2$D(1$_{11}$--1$_{01}$), NH$_3$(1,1), N$_2$D$^+$(1--0) or N$_2$D$^+$(2--1) and N$_2$H$^+$(1--0) towards the observed cores. The colored dots show the integrated intensity peaks and contours show 60\% of the integrated intensity peak. The beam size is shown in the bottom left corner of each map. The black cross shows the positions observed in  \citet{Punanova2016}.}
\label{Contr}
\end{figure*}

\begin{table}
\begin{center}
\begin{tabular}{| l | c  | c  |}\hline
Core & $\alpha$(J2000) &$\delta$(J2000) \\ \hline
Oph-C-N & 16$^h$26$^m$57.2$^s$  & -24$^{\circ}$31$^{\prime}$39.0$^{\prime\prime}$  \\ 
Oph-E-MM2 &16$^h$27$^m$03.1$^s$  &-24$^{\circ}$39$^{\prime}$15.0$^{\prime\prime}$  \\ 
Oph-F & 16$^h$27$^m$24.7$^s$  & -24$^{\circ}$40$^{\prime}$35.3$^{\prime\prime}$ \\ 
Oph-H-MM1 &16$^h$27$^m$58.3$^s$  & -24$^{\circ}$33$^{\prime}$43.4$^{\prime\prime}$  \\ 
Oph-H-MM2* &  16$^h$28$^m$20.7$^s$ & -24$^{\circ}$36$^{\prime}$21.3$^{\prime\prime}$ \\
Oph-I-MM2* &  16$^h$28$^m$33.1$^s$ & -24$^{\circ}$18$^{\prime}$28.4$^{\prime\prime}$ \\
\hline
\end{tabular}
\end{center}
\caption{The coordinates of the dust emission peaks of the studied cold dense cores \citep{DiFrancesco2008} and of the ammonia peaks \citep[with asterisk*,][]{Frisen2017}.}\label{tab:coord}
\end{table}

\begin{table*}
\begin{center}
\begin{tabular}{| l | l | c | c | c | c | c | c | c | c | c | c | c |}
\hline
Core &Transition & Frequency &$T_{\rm sys}$ & $HPBW_{\rm obs}$ & rms, $T_{\rm mb}$ & $F_{\rm eff}$ & $B_{\rm eff}$ & $\Delta v_{\rm res}$ &   $T_{\rm ex}$  \\ 
 & & (GHz) & (K) & ($^{\prime\prime}$) & (K) & & & (km s$^{-1}$) &  (K)\\ \hline
Oph-C-N& N$_2$H$^+$(1--0) & ~93.1737637$^a$   &    143     &  26.4$^{\prime\prime}$ & 0.08 - 0.10  &0.95 & 0.80        &          0.063          &  6.0$\pm$0.6\\
&N$_2$D$^+$(1--0)& ~77.1096162$^a$  &    173     &  33.6$^{\prime\prime}$ & 0.10  &0.95  & 0.81       &          0.075          &  4.5$\pm$0.5 \\
 & $p$-NH$_2$D(1$_{11}$--1$_{01}$) & 110.153599$^b$   &   321     &  23.3$^{\prime\prime}$& 0.05 &0.95   & 0.79     &           0.053     &       4.7$\pm$0.9 \\
&H$^{13}$CO$^+$(2--1)& 173.5067003$^c$  &  1542  &  14.2$^{\prime\prime}$& 0.61  &0.93  & 0.69     &            0.134       &    10.0  \\
&DCO$^+$(2--1)& ~144.0772804$^d$  &  227    &  17.1$^{\prime\prime}$& 0.06  &0.93        & 0.73  &       0.041        &  7.0$\pm$0.6   \\ \hline

Oph-E-MM2& N$_2$H$^+$(1--0)* & ~93.1737637$^a$   &    --     &  26.4$^{\prime\prime}$ & 0.19  & &      &          0.063          &  6.0$\pm$0.7\\
&N$_2$D$^+$(1--0) & ~77.1096162$^a$   &    156   &  33.6$^{\prime\prime}$ & 0.12  &0.95 & 0.81      &            0.076         &   4.9$\pm$0.5  \\
&$p$-NH$_2$D(1$_{11}$--1$_{01}$)& 110.153599$^b$   &     265   &   23.5$^{\prime\prime}$& 0.06   &0.95  & 0.79     &            0.053       &    4.5$\pm$0.5  \\
&H$^{13}$CO$^+$(1--0)& ~86.7542884$^c$   &  207    &   28.4$^{\prime\prime}$& 0.10   &0.95  & 0.80      &          0.067         &  10.0   \\
&DCO$^+$(2--1)& 144.0772804$^d$  &  246     &   17.1$^{\prime\prime}$& 0.05   &0.93  &  0.73     &          0.041        &   8.0$\pm$0.7  \\ \hline

Oph-F& N$_2$H$^+$(1--0)* & ~93.1737637$^a$   &    --     &  26.4$^{\prime\prime}$ & 0.16  & &      &          0.063           &  7.0$\pm$1.1\\
&N$_2$D$^+$(2--1) & 154.2171805$^a$  & 366   &   16.0$^{\prime\prime}$& 0.08   &0.95 & 0.72      &          0.076       &      5.0$\pm$0.9\\
&$p$-NH$_2$D(1$_{11}$--1$_{01}$)& 110.153599$^b$    &    265   &   22.5$^{\prime\prime}$& 0.10   &0.95  & 0.79         &      0.053          &   4.9$\pm$0.8  \\ \hline

Oph-H-MM1&N$_2$H$^+$(1--0) & 93.1737637$^a$   &     133   &  26.4$^{\prime\prime}$& 0.09   &0.95 & 0.80     &       0.063          &   6.5$\pm$0.6   \\
&N$_2$D$^+$(1--0) & ~77.1096162$^a$   &     189  &  33.6$^{\prime\prime}$& 0.09   & 0.95 & 0.81      &         0.076       &     4.2 $\pm$0.6  \\ \hline
Oph-H-MM2**& N$_2$H$^+$(1--0)  & 93.1737637$^a$ & 176 & 26.4$^{\prime\prime}$  & 0.08 & 0.95 & 0.80 & 0.063 & 7.0$\pm$1.1\\
& N$_2$D$^+$(1--0) & 77.1096162$^a$ & 177 & 33.6$^{\prime\prime}$  & 0.13 & 0.95 & 0.81 & 0.076 & 4.7$\pm$0.6\\
& N$_2$D$^+$(2--1) & 154.2171805$^a$ & 325 & 16.0$^{\prime\prime}$  & 0.10 & 0.93 & 0.72 & 0.038 & 5.0$\pm$0.7 \\
& DCO$^+$(2--1) & 144.0772804$^d$ & 229 & 17.1$^{\prime\prime}$  & 0.12 & 0.93 & 0.73 & 0.041 &  6.9$\pm$0.9\\ \hline
Oph-I-MM2** & N$_2$H$^+$(1--0) & 93.1737637$^a$ & 180 & 26.4$^{\prime\prime}$ & 0.07 & 0.95 & 0.80 & 0.063 & 6.5$\pm$0.9\\
& N$_2$D$^+$(1--0) & 77.1096162$^a$ & 236 & 33.6$^{\prime\prime}$ & 0.11 & 0.95 &  0.81 & 0.076 & 4.5$\pm$0.6\\
& N$_2$D$^+$(2--1) & 154.2171805$^a$ & 337 & 16.0$^{\prime\prime}$ & 0.08  & 0.93 & 0.72 & 0.038 &  5.0$\pm$0.8\\
& DCO$^+$(2--1) & 144.0772804$^d$ & 368 & 17.1$^{\prime\prime}$ & 0.10 & 0.93 & 0.73 & 0.041 & 6.7$\pm$1.1\\ \hline
\end{tabular}
\end{center}
\begin{flushleft}
      \small
      \begin{flushleft}
      \item \textbf{Note:} $HPBW_{\rm obs}$ is the native half-power beam width at the given frequency. The transition frequencies, the frequencies and relative intensities of the hyperfine structure components are taken from the following works: $^a$ from \cite{Pagani2009_N2H_N2D}, $^b$ from \cite{Daniel2016}, $^c$ from \cite{Schmid-Burgk2004_H13CO}, $^d$ from \cite{Lattanzi2007_DCO21}. The frequencies and relative intensities of the hyperfine structure components are also available via CDMS \citep{CDMS}. *N$_2$H$^+$(1--0) data for Oph-E-MM2 and Oph-F are taken from \citet{Andre2007}. **Cores Oph-H-MM2 and Oph-I-MM2 are observed only with single pointings. The high system temperature and high noise level of the H$^{13}$CO$^+$(2--1) line is due to the strong atmospheric water absorption feature centered at 183.3 GHz.
      \end{flushleft}
\end{flushleft}

\caption{The observed lines, observational parameters and the excitation temperature used for the fits.}\label{tab:allline}
\end{table*}

\section{Data reduction and analysis}\label{sec:data_reduction}

\subsection{Spectral maps}
We convolved all maps to the same angular resolution of 33.6$^{\prime\prime}$, corresponding to the largest beam size in our data set. The pixel size was 12$^{\prime\prime}$, consistent with the Nyquist criterion. The map of H$^{13}$CO$^+$(2--1) for Oph-C-N was smoothed to the velocity resolution of 0.134 km~s$^{-1}$ to increase the signal-to-nose ratio. Up to the stage of spectral cubes, the spectra were processed with GILDAS\footnote{ GILDAS: \url{https://www.iram.fr/IRAMFR/GILDAS/}} software.

\subsection{Line analysis}\label{sect:line_analysis}
For line analysis, we took the parts of the spectral cubes with the uniform noise (shown in Table~\ref{tab:allline}) and signal-to-noise ratio $>4$. All observed transitions have hyperfine structure (hfs). We fitted the spectra to find the line parameters using the python package pyspeckit\footnote{Pyspeckit: \url{https://github.com/pyspeckit/pyspeckit}} \citep{Ginsburg2011,Ginsburg2022}. The package fits the hyperfine structures using the frequencies and the relative intensities of the hfs components under the assumption of local thermodynamic equilibrium. The fitted spectral parameters are the transition excitation temperature $T_{\rm ex}$, optical depth $\tau$, radial velocity $V_{\rm LSR}$, and the velocity dispersion $\sigma$ (connected to the full width at half maximum, FWHM, as $\sigma$ = $FWHM$/$2\sqrt{2\cdot\ln{2}}$). The package uses the gradient descent method and the radiation transfer equation to determine the transition excitation temperature and optical depth.

For each parameter map, pyspeckit creates a parameter uncertainty map. 
Estimations show that the radial velocity and line width have uncertainties between 0.001--0.010~km~s$^{-1}$. We corrected velocity dispersion for the channel width:
\begin{equation}
\sigma = \sqrt{\sigma _{\rm obs}^2-\Delta v_{\rm res}^2/(8 \ln{2})} ,
\end{equation}
where $\sigma _{\rm obs}$ -- observed velocity dispersion, $\Delta v_{\rm res}$ -- channel width. In the majority of the pixels in all maps except for NH$_3$, $\tau$ and $T_{\rm ex}$ had large uncertainties, 30--100\% of the values. To avoid such large uncertainties, we fix the parameter that affects the resulting column density the least. We explored the parameter space to find any correlation between the parameters. We found strong correlation between optical depth and excitation temperature (see Fig.~\ref{pic: zone1}).
The spread of excitation temperature values has a smaller range (about 4~K) than that of the optical depth (up to 10). The variation of excitation temperature affects the column density less than the variation of optical depth, which affects the column density linearly: at the temperatures of 4--8~K, the column density increases with temperature by 25--30\% per 1~K for N$_2$H$^+$ and N$_2$D$^+$, by 12--15\% per 1~K for DCO$^+$, and by 2--5\% per 1~K for NH$_2$D. Thus we decided to use a single excitation temperature value for each line towards each core to estimate the optical depth and get a better estimate of the column density. To find the most suitable excitation temperature value, we explored the parameter space of the optical depth and the excitation temperature and produced the probability field with the Monte-Carlo method (see details in Appendix~\ref{depth}).

We show the $T_{\rm ex}$ found via parameter space exploration in the last column of Table~\ref{tab:allline}. The uncertainty of the excitation temperature was estimated as the standard deviation of the average excitation temperature in the most probable area and amounts to 10--20\%. We could not find the most probable excitation temperature for the H$^{13}$CO$^+$(1--0) line. We considered the H$^{13}$CO$^+$(1--0) line optically thin, fixed $\tau = 0.1$, and used $T_{\rm ex}$=10~K to determine the rest of the parameters as was done in \citet{Punanova2016}. Since the H$^{13}$CO$^+$(2--1) line is very narrow and unsaturated, we also considered it optically thin and fitted with a Gaussian. The optical depth $\tau$ is $> 0.1$ in the maps for all other transitions. 

To estimate the goodness of the fit, we subtracted the resulting spectrum model from the observed spectrum. The rms of the subtraction does not exceed the rms noise level of the observed spectrum (the difference is within 10\%). 
Therefore, the model spectra with the fixed excitation temperature were sufficient. The spectra with the highest signal-to-noise ratio and their fits for each spectral cube (for each core and tracer) are shown in Fig.~\ref{pic: specCN}.

For NH$_3$ hfs line analysis, we used the method described in \cite{Rosolowsky_2008,Pyspeackit2011}. We used the observational maps of para-NH$_3$(1,1) and para-NH$_3$(2,2) from the Green Bank ammonia survey \citep[GAS,][]{Frisen2017} to obtain the NH$_3$ column density. 
The spectral parameters were fitted with good accuracy (median relative discrepancy between the model and the spectra = 13\%), since two spectral transitions are used in the fitting. The method `fiteach' takes into account the $o/p$ ratio for NH$_3$ as a parameter, which we set to $o/p$~=~1:1 following \citet{Harju2017}, and gives the total column density of all (ortho + para) NH$_3$. 

\section{Results}\label{sec:results}

\subsection{The distribution of emission}
Figure~\ref{Contr} shows the 60\% contours of the peak integrated intensity of the observed transitions that represent the distribution of the observed tracers. The colored dots show the integrated intensity peaks. In Oph-C-N, where we have all six tracers observed, we see that the emission of the carbon-bearing DCO$^+$ shows the most extended spatial distribution. Among the nitrogen-bearing species, deuterated isotopologues (N$_2$D$^+$ and NH$_2$D) show more compact emission compared to those with hydrogen. NH$_2$D shows the most compact emission and appears to be the best tracer of dense gas in our selection of lines.
H$^{13}$CO$^+$(2--1) shows a very small emission area because of the low signal-to-noise level of the spectra. The integrated intensity maps are presented in Fig.~\ref{pic: W_map}.

\subsection{Spectrum parameters}\label{sec:spparam}
Figures \ref{pic: tau}, \ref{pic: vel} and \ref{pic: sigma} show the resulting fit parameters: the maps of the optical depth $\tau$ (Fig.~\ref{pic: tau}, except for NH$_3$ where the fit gives directly $N_{\rm tot}$), the radial velocities $V_{\rm LSR}$ (Fig.~\ref{pic: vel}), and the velocity dispersion $\sigma$ (Fig.~\ref{pic: sigma}) are presented for each transition line in each core. The used excitation temperature is given in the last column of Table~\ref{tab:allline}. 
The parameter maps are masked to show only the values with the relative uncertainty smaller than $1/3$.

\subsubsection{Velocity dispersions}

\begin{figure*}
\centering
\includegraphics[scale=0.33]{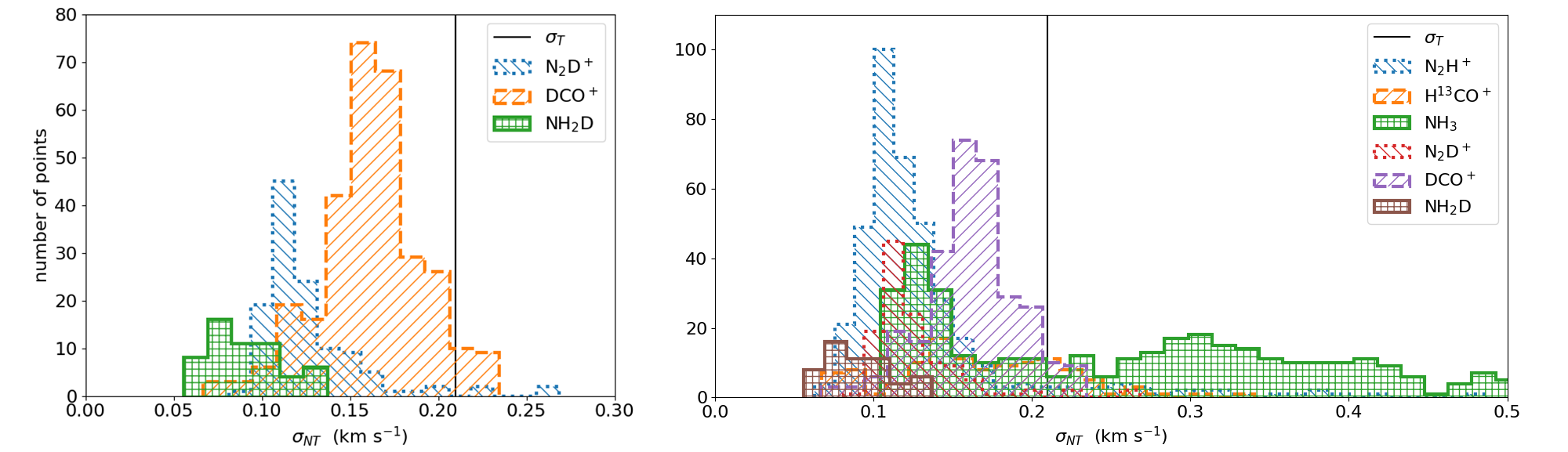}
\caption{Histogram of the non-thermal component of the velocity dispersion $\sigma_{\rm NT}$ of deuterated species (left) and for all species (right) for all cores. The step is 0.011~km~s$^{-1}$. The black line shows the thermal velocity dispersion $\sigma_T$ at 12 K. }
\label{pic:hist}
\end{figure*}

The velocity dispersion for all lines in the studied dense cores ranges from 0.05~km~s$^{-1}$ to 1.00~km~s$^{-1}$ (see Fig.~\ref{pic: sigma}). The maximum velocity dispersions of 0.85 and 1.00~km~s$^{-1}$ are observed in N$_2$H$^+$ and NH$_3$ towards Oph-F core  (row 3, columns 3 and 6 in Fig.~\ref{pic: sigma}). The rest of the species show moderate and low velocity dispersions, up to 0.34~km~s$^{-1}$, with median values of 0.07--0.17~km~s$^{-1}$. The lowest velocity dispersion are observed in NH$_2$D, 0.05--0.14~km~s$^{-1}$ with a median value of 0.10~km~s$^{-1}$ (see the third column in Fig.~\ref{pic: sigma}).

The general picture of narrow lines ($\sigma$=0.10--0.15~km~s$^{-1}$) is sometimes disturbed with spots of larger velocity dispersion ($\sigma$=0.20--0.30~km~s$^{-1}$). Visual inspection of the spectra towards the north-eastern side of Oph-C-N (see Fig.~\ref{pic: sigma}, row 1), the strip of increased velocity dispersion as seen in all lines, reveals the presence of a second velocity component. The velocities of the two peaks differ by 0.3--0.4~km~s$^{-1}$. The north-western part of Oph-F (see Fig.~\ref{pic: sigma}, row 3) also shows an increased velocity dispersions in N$_2$H$^+$ and N$_2$D$^+$. The N$_2$H$^+$(1--0) lines show a second velocity component there \citep[also reported in][]{Andre2007,Punanova2016}. Additional components that affect the overall velocity field were also found towards the Oph-H-MM1 core in \citet{Pineda2022}. However, the study of gas kinematics is out of scope of this paper, thus we do not analyse the two velocity components.
The cores Oph-E-MM2, Oph-H-MM1, Oph-H-MM2, Oph-I-MM2 do not show the presence of the second velocity components in our data. 

Ammonia maps (the far right column in Fig.~\ref{pic: sigma}) show up to $0.07$~km~s$^{-1}$ larger velocity dispersion compared to the other N-bearing species within the cores (the three left columns in Fig.~\ref{pic: sigma}), and, outside of the cores, up to $0.7$~km~s$^{-1}$ larger velocity dispersion compared to all other species, if we define the cores as the area with subsonic velocity dispersion \citep[as in, e.g.,][]{Pineda2010,Chen2019}. The large dispersions are due to the contribution of the surrounding cloud velocity component \citep{Choudhury2020} which has not been taken into account in our analysis. However, towards the cores, the narrow lines emitted by the core dense gas dominate the emission, and the NH$_3$ velocity dispersion is similar to that of the other species (see Fig.~\ref{pic: sigma}) and shows stronger decrease towards the core centres than, e.g., N$_2$H$^+$ (see the ratio of the described below non-thermal components in Fig.~\ref{pic:Sigma_Pineda}), similar to the result of \citet{Pineda2021}, who found that, in dense gas, NH$_3$ shows smaller velocity dispersions than N$_2$H$^+$. 

We estimated the non-thermal component of velocity dispersion as follows \citep{Myers1991}:
\begin{equation}
\Delta \sigma_{\rm NT}^2 = \Delta \sigma_{\rm obs}^2 - \frac{k T_k}{m_{\rm obs}},
\end{equation}
where $k$ is the Boltzmann’s constant, $T_k$ is the gas kinetic temperature taken from \citet{Frisen2017}, and $m_{\rm obs}$ is the mass of the observed molecule in a.m.u.. The thermal component is 0.053--0.070~km~s$^{-1}$ for the selected species. The velocity dispersion in all cores is low (median dispersions are 0.07--0.17~km~s$^{-1}$), which indicates mostly thermal or subthermal motions, that is, low turbulence.

Figure~\ref{pic:hist} compares the non-thermal components of velocity dispersions of all species (right panel) with the thermal velocity dispersion (black line) at 12~K, which is the median gas temperature in the studied cores. We focus on the emission of the deuterated species, N$_2$D$^+$, DCO$^+$ and NH$_2$D as the least contaminated by the emission of the surrounding cloud gas (left panel). The velocity dispersion in the majority of the data points is subthermal. The velocity dispersion in DCO$^+$, $\sim$0.16 km~s$^{-1}$ (the shells of the cores), is greater than that in N$_2$D$^+$ and NH$_2$D $\sim$0.11 km~s$^{-1}$ and $\sim$0.08 km~s$^{-1}$ (the centers of the cores). NH$_2$D shows the narrowest lines which may indicate that they originate in even colder gas in the core centers.

\subsection{Column density}\label{sec:colden}

\begin{table*}
\begin{center}
\begin{tabular}{| l | c  | c  |c  |c  |c  |c l}\hline
Transition & $A_{\rm ul}$ & $Q_{\rm rot}$ & $E_l$ & $g_l$ & $g_u$ & $n_c^{10 {\rm K}}$ & Citation for rotational constants, dipole moments *, and collisional rates **\\ 
 & (10$^{-5}$ s$^{-1}$) &   &  (K)& & & (10$^4$ cm$^{-3}$)\\ \hline
N$_2$H$^+$(1--0) & ~3.628 &3.043  & 0&1&3 & 6.1 & *\citet{Cazzoli2012,AMANO2005}\\ 
N$_2$D$^+$(1--0) & ~2.056& 2.794 &  0&1&3 & 6.2 & *\citet{Dore2004,AMANO2005}, **\citet{Dore2004}\\ 
N$_2$D$^+$(2--1) &16.450 & 3.060  & ~7.41&3&5 & 55\\ 
$p$-NH$_3$(1,1) & ~0.014 & 4.301 &  28.64&3&3 & 0.2 & *\citet{Yu2010}\\ 
$p$-NH$_2$D(1$_{11}$--1$_{01}$) & ~1.650 & 27.110 &  15.97&3&3 & 7.3 & *\citet{Fusina1988_NH2D,COHEN1982}, **\citet{Daniel2014} \\ 
H$^{13}$CO$^+$(1--0)  & ~3.853  & 4.970 &0&1&3 & 6.2 & *\citet{Schmid-Burgk2004_H13CO,Lattanzi2007}\\
H$^{13}$CO$^+$(2--1)  & 36.825 & 5.149 & ~4.17&3&5 & 51\\
DCO$^+$(2--1) &13.414 &4.970 & ~3.46&3&5 & 71 & *\citet{Lattanzi2007_DCO21,vanderTak2009}, **\citet{Denis-Alpizar2020}\\
\hline
\end{tabular}
\end{center}
\begin{flushleft}
      \small
      \begin{flushleft}
      \item \textbf{Note:} The partition functions $Q_{\rm rot}$, lower level energies $E_l$, and Einstein coefficients $A_{\rm ul}$ of the linear molecules were calculated as in \citet{Caselli2002}. For that, the rotational constants and dipole moments are taken from the works, listed in the right column* 
      and are also available through the CDMS data base \citep{CDMS}. The Einstein coefficient ($A_{\rm ul}$) and the lower level energy ($E_l$) for $p$-NH$_2$D(1$_{11}$--1$_{01}$) were taken from \citet{Harju2017}. $Q_{\rm rot}$ for $p$-NH$_2$D(1$_{11}$--1$_{01}$) was taken from the CDMS data base \citep{CDMS}. The critical densities ($n_c^{10 {\rm K}}$) of hydrogenated species were taken from \citet{Yancy2015} and those of deuterated species were calculated in the same manner as in \citet{Yancy2015} using the collisional rates from the works listed in the right column** and available through the LAMBDA data base \citep{Schoiler2005}. 
      
      \end{flushleft}
\end{flushleft}

\caption{Quantum coefficients used to calculate the column densities.  }\label{tab:cof}
\end{table*}

\begin{figure*}
\centering
\includegraphics[scale=0.38]{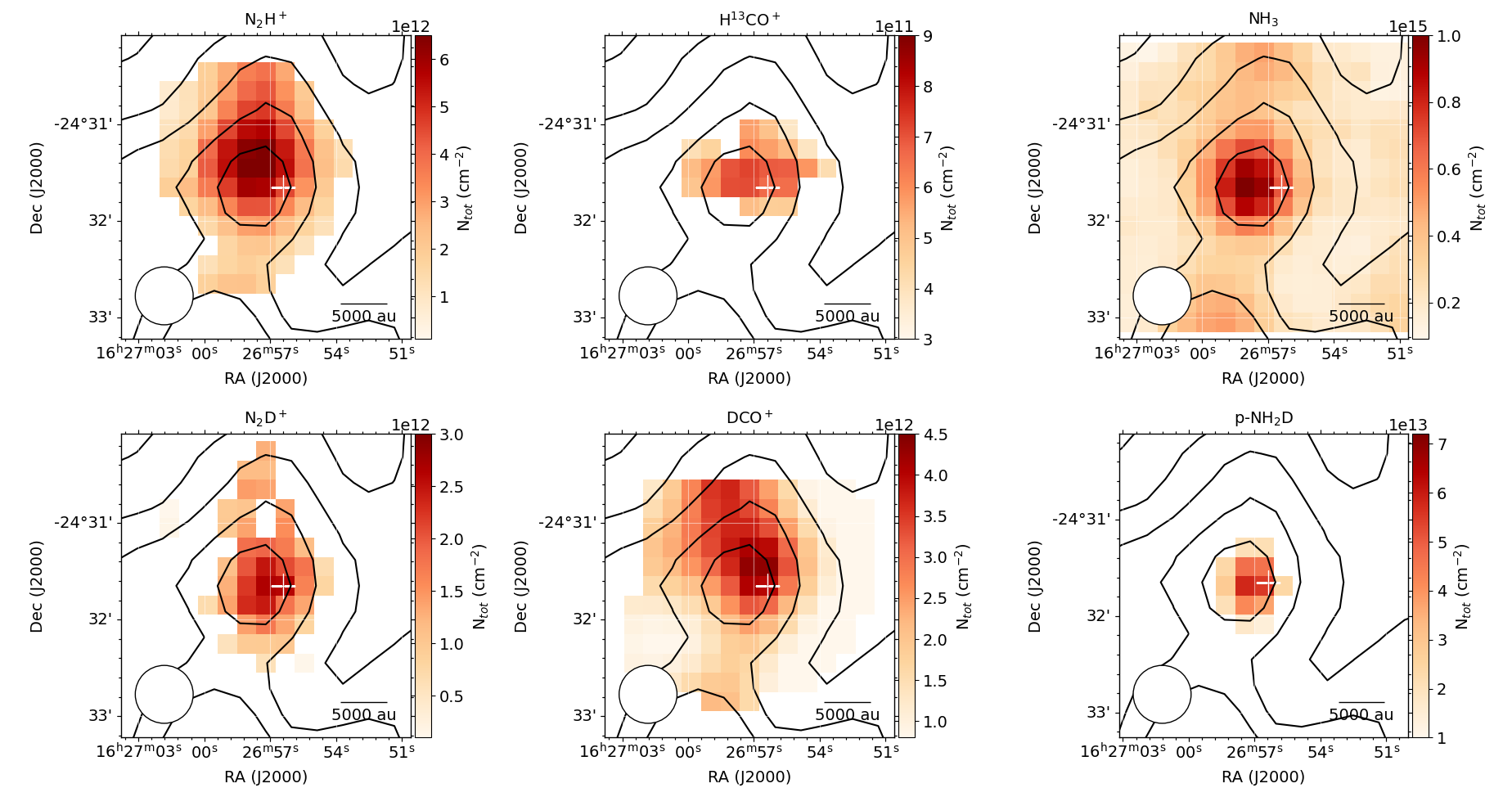}
\caption{The column density maps of N$_2$H$^+$, N$_2$D$^+$, $p$-NH$_2$D, NH$_3$, DCO$^+$, and  H$^{13}$CO$^+$ towards Oph-C-N. The molecular hydrogen column density is shown by black contours. The first contour starts at 1.5$\times$10$^{22}$~cm$^{-2}$ with a contour step of 0.5$\times$10$^{22}$~cm$^{-2}$. The beam size is shown in the bottom left corner of each map. The white cross shows the position observed in  \citet{Punanova2016}.}

\label{pic:Oph-C-N_N}
\end{figure*}

\begin{figure*}
\centering
\includegraphics[scale=0.38]{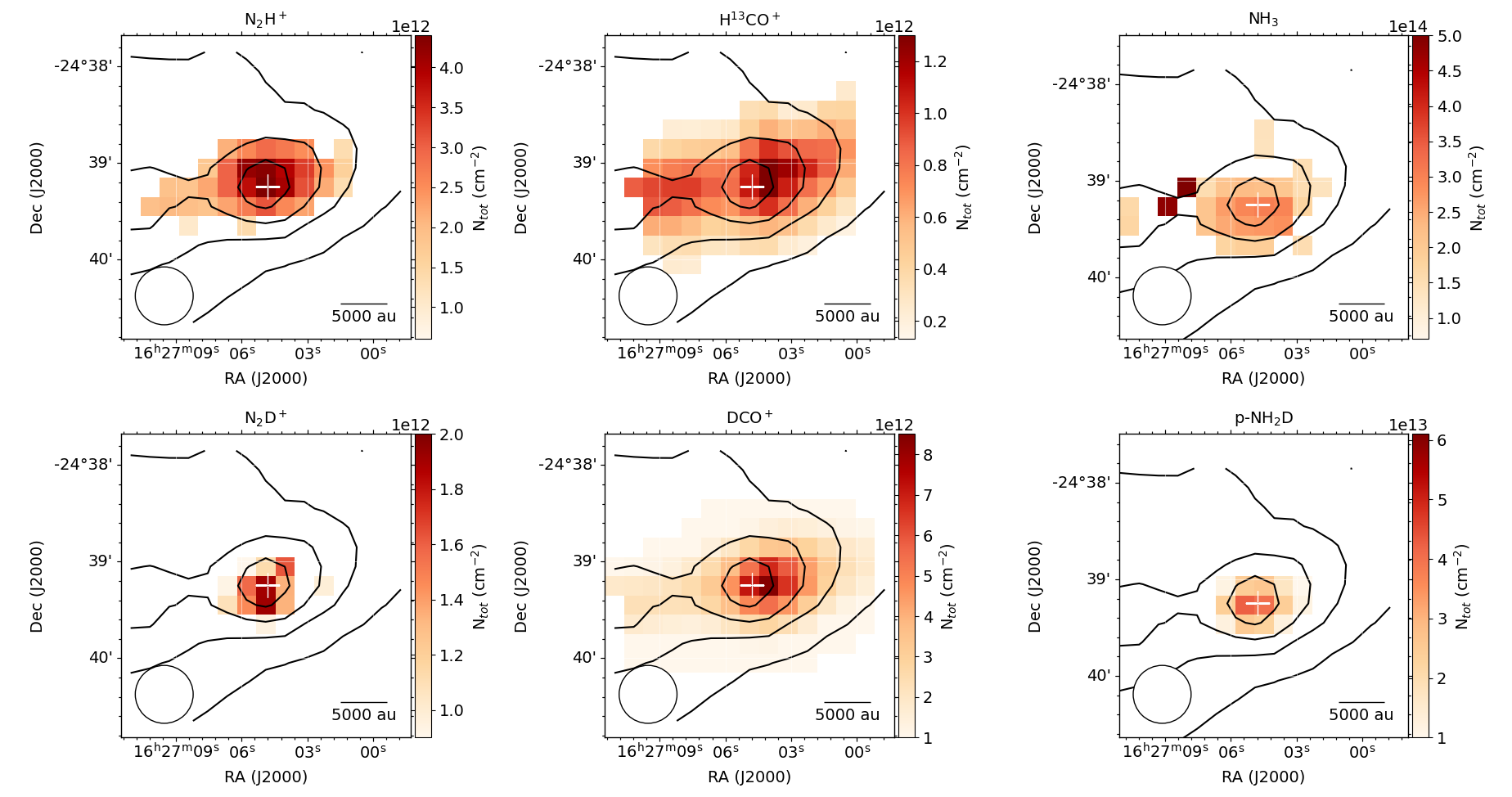}
\caption{The column density maps of N$_2$H$^+$, N$_2$D$^+$, $p$-NH$_2$D, NH$_3$, DCO$^+$, and  H$^{13}$CO$^+$ towards Oph-E-MM2. The molecular hydrogen column density is shown by black contours. The first contour starts at 1.5$\times$10$^{22}$~cm$^{-2}$ with a contour step of 0.5$\times$10$^{22}$~cm$^{-2}$. The beam size is shown in the bottom left corner of each map. The white cross shows the position observed in  \citet{Punanova2016}.}

\label{pic:Oph-E-MM2_N}
\end{figure*}

\begin{figure*}
\centering
\includegraphics[scale=0.38]{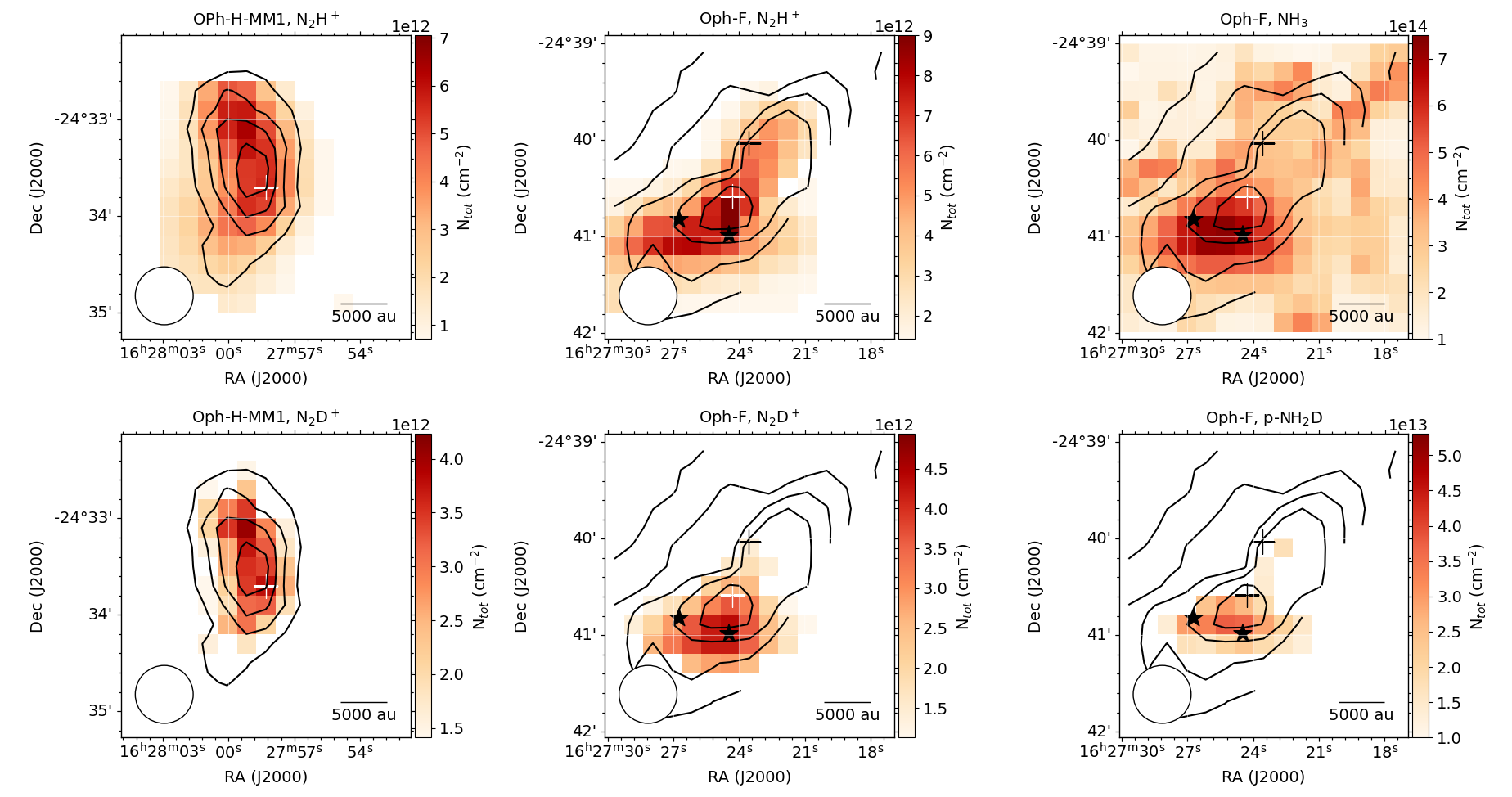}
\caption{The column density maps of N$_2$H$^+$, N$_2$D$^+$ towards Oph-H-MM1 and the column density maps of N$_2$H$^+$, N$_2$D$^+$, $p$-NH$_2$D, and NH$_3$  towards Oph-F. The molecular hydrogen column density is shown by black contours. The first contour starts at  1.5$\times$10$^{22}$~cm$^{-2}$  with a contour step of 0.5$\times$10$^{22}$~cm$^{-2}$. The beam size is shown in the bottom left corner of each map. The white crosses show the positions observed in \citet{Punanova2016}. The stars show the positions of the YSOs \citep[YLW15, Class~0$+$I, CRBR~2422.8-3423, Class~I;][]{Young1986,Kirk2017,Comeron1993,Bontemps2001} in Oph-F.}

\label{pic:Oph-F_N}
\end{figure*}

\begin{table*}
\begin{center}
\begin{tabular}{| l | l | c | c | c | c | c | c | c |}
\hline
Core & Line  & $T_{\rm ex}$ & $\tau$ & $V_{\rm LSR}$ & $\sigma$& $N_{\rm tot}$ & $R_D$ \\
      &           &  (K)   &         & (km~s$^{-1}$) & (km~s$^{-1}$) & (10$^{12}$~cm$^{-2}$) & \\ \hline
Oph-H-MM2           &  DCO$^+$(2--1)  &  6.9 & 4.6$\pm$2.1 & 4.081$\pm$0.008 & 0.115$\pm$0.014 & 2.83$\pm$1.32 & -- \\ 
                    & N$_2$H$^+$(1--0)   & 7.0 & 7.7$\pm$0.6 &  4.096$\pm$0.005 & 0.096$\pm$0.005 & 5.09$\pm$0.74 & --\\
                    & N$_2$D$^+$(1--0)    & 3.5 &  2.7$\pm$0.3 & 4.065$\pm$0.011 & 0.110$\pm$0.011 & 1.02$\pm$0.25 & 0.20$\pm$0.08  \\
                    & N$_2$D$^+$(2--1)  & 5.0 & 2.1$\pm$0.4  & 4.061$\pm$0.021 & 0.104$\pm$0.017 & 1.33$\pm$0.26 & 0.26$\pm$0.09 \\
Oph-I-MM2             & DCO$^+$(2--1)  & 7.0 & 3.8$\pm$0.8& 3.331$\pm$0.008& 0.095$\pm$0.015 & 1.74$\pm$0.39 & -- \\
                    & N$_2$H$^+$(1--0)   & 6.8 & 5.3$\pm$0.3 & 3.331$\pm$0.004 & 0.079$\pm$0.006 & 2.18$\pm$0.37 & -- \\
                    & N$_2$D$^+$(1--0)    &   4.5   &  1.5$\pm$0.1    &  3.332$\pm$0.009  &   0.101$\pm$0.009 &  0.69$\pm$0.14 & 0.32$\pm$0.08 \\
                    &   N$_2$D$^+$(2--1)  & 5.0 & 3.2$\pm$0.5 & 3.311$\pm$0.008 & 0.054$\pm$0.007 & 0.83$\pm$.0.21&  0.38$\pm$0.11 \\ 
\hline
\end{tabular}
\end{center}
\caption{Line parameters, column densities, and deuterium fractions towards Oph-H-MM2 and Oph-I-MM2.} 
\label{tab:single-point}
\end{table*}

The formula to calculate the column density can be obtained from the radiation transfer equation. For this it is necessary to write down the intensities of the rotational transition and the absorption coefficient in terms of the number of molecules. We assume that cold dense cores are in local thermodynamic equilibrium \citep[e.g.,][]{Shirley2015}. Column density of NH$_3$ was obtained from the fit; for the rest of the species (except for H$^{13}$CO$^+$) we calculate the column density as it was done in \citet{Caselli2002} for optically thick transitions:
\begin{equation}
N_{\rm tot}=\frac{8\pi^{3/2}\sigma g_l}{\lambda^3A_{\rm ul}g_u} \frac{\tau}{1-\exp(-h\nu/kT_{\rm ex})} \frac{Q_{\rm rot}}{g_l\exp(-E_l/kT_{\rm ex})},
\end{equation}
where $\sigma$ is the velocity dispersion, $g_l$ and $g_u$ are the statistical weights of the lower and upper energy levels,  $\lambda$ is the transition wavelength, $A_{\rm ul}$ is the Einstein coefficient, $\tau$ is the optical depth, $h$~is the Planck constant, $\nu$~is the transition frequency, $k$ is the Boltzmann constant, $T_{\rm ex}$~is the excitation temperature, $Q_{\rm rot}$ is the partition function, $E_l$ is the energy of the lower level. The partition function of a linear molecule (N$_2$H$^+$, N$_2$D$^+$, H$^{13}$CO$^+$, DCO$^+$) can be calculated as follows:
\begin{equation}
Q_{\rm rot}=\sum^\infty_{J=0}(2J+1)\exp(E_J/kT),
\end{equation}
where $J$ is the rotational quantum number, $E_J$ -- rotational transition energy ($J(J+1)hB$), $T$ is the gas temperature.
The partition function for deuterated ammonia was taken from the CDMS data base \citep{CDMS}. All quantum coefficients used to calculate the column densities are listed in Table~\ref{tab:cof}.

Since we assumed that the H$^{13}$CO$^+$(1--0) and (2--1) lines are optically thin, the column density of H$^{13}$CO$^+$ was calculated as in \citet{Caselli2002} for optically thin transitions:

\begin{multline}
N_{\rm tot} =\frac{8W\pi g_l}{\lambda^3A_{\rm ul}g_u} \frac{1}{J_\nu(T_{\rm ex})-J_\nu(T_{\rm bg})}\times\\  \times \frac{1}{1-\exp(-hv/kT_{\rm ex})}\frac{Q_{\rm rot}}{\exp(-E/kT_{\rm ex})},
\end{multline}
where  $W$ is the integrated intensity, $J_\nu(T)$ is the equivalent Rayleigh-Jeans temperature, $T_{\rm bg}$ is the background temperature of 2.7~K. 

The column density of ammonia is calculated using the pyspeckit fitter (see Sect.~\ref{sec:data_reduction}). The $o/p$ ratio of (NH$_3$) is introduced in the fitter, we set it to 1:1 (the last parameter of the ammonia fitter is fixed to 0.5) following \citet{Harju2017}. 

\begin{table*}
\begin{center}
\begin{tabular}{lccccc||cccccc}\hline
 & \multicolumn{5}{c||}{This work} & &\multicolumn{5}{c}{In \citet{Punanova2016}} \\ \cmidrule{2-6} \cmidrule{8-12}
Core 	& $N_{\rm tot}$(N$_2$H$^+$) 	& $N_{\rm tot}$(N$_2$D$^+$) & $T_{\rm ex}$* & $\tau$*	 & $R_D$ 	& & $N_{\rm tot}$(N$_2$H$^+$) & $N_{\rm tot}$(N$_2$D$^+$)  & $T_{\rm ex}$* & $\tau$*  & $R_D$\\ 
	& (10$^{13}$ cm$^{-2}$) 	& (10$^{13}$ cm$^{-2}$)	    & (K)	   &	&	& & (10$^{13}$ cm$^{-2}$)     & (10$^{13}$ cm$^{-2}$) & (K) & & \\ \hline

Oph-C-N & 0.45$\pm$0.06 		& 0.25$\pm$0.03  & 6.0/4.5 &	6.8/3.7   & 0.53$\pm$0.14	      		& & 1.23$\pm$0.15	     & 0.53$\pm$0.11 & 6.5/4.4	& 12.3/6.4	    & 0.43$\pm$0.10			\\ 
Oph-E-MM2 & 0.44$\pm$0.06		& 0.20$\pm$0.02  & 6.0/4.9 &4.3/1.7  	 & 0.43$\pm$0.11	      		& & 0.59$\pm$0.17	     & 0.45$\pm$0.13 & 8.2/4.0	& 2.8/4.9	    & 0.42$\pm$0.02			\\ 
Oph-F 	&0.89$\pm$0.13 			&  0.27$\pm$0.04 & 7.0/5.0 &1.9/1.2  	 & 0.41$\pm$0.13	      		& & 1.18$\pm$0.15	     & 0.15$\pm$0.02 & 5.4/5.4	& 8.8/0.1	    & 0.13$\pm$0.02			\\ 
Oph-H-MM1 & 0.56$\pm$0.07 		& 0.31$\pm$0.06  & 6.5/4.2 &	4.7/4.8 & 0.67$\pm$0.17	      		& & 0.88$\pm$0.09	     & 0.38$\pm$0.09 & 9.3/5.5	& 3.7/2.7	    & 0.43$\pm$0.11			\\ 
\hline
\end{tabular}
\end{center}
\caption{The comparison of our column densities and deuterium fractions (left part of the table) with those from \citet[][right part of the table]{Punanova2016}. For Oph-F we show the values of F-MM2 in \citet{Punanova2016}. *The $T_{\rm ex}$ and $\tau$ are separated with a slash for N$_2$H$^+$ and N$_2$D$^+$.}\label{tab:N}
\end{table*}

Figures~\ref{pic:Oph-C-N_N}, \ref{pic:Oph-E-MM2_N}, and \ref{pic:Oph-F_N} show the column density maps of all species and Table \ref{tab:single-point} shows the column density for the single-pointing observations towards Oph-H-MM2 and Oph-I-MM2. In Oph-C-N and Oph-E-MM2, the column density of all species increases towards the dust peaks in cores. In Oph-F, the molecular emission peaks are slightly shifted to the south-east from the dust peak by one to half beamsize. In Oph-H-MM1, there are two molecular peaks \citep[the two peaks apper prominently in the interferometric view of NH$_3$;][]{Pineda2022}, and N$_2$D$^+$ shows strong emission across all the core. The column density is $\sim$10$^{12}$~cm$^{-2}$ in N$_2$H$^+$, N$_2$D$^+$, $\sim$10$^{13}$~cm$^{-2}$ in $p$-NH$_2$D, and $\sim$10$^{14}$~cm$^{-2}$ in NH$_3$ for all cores. In H$^{13}$CO$^+$ and DCO$^+$, column densities are  $\sim$10$^{11-12}$~cm$^{-2}$ for Oph-E-MM2 and Oph-C-N.

In Table~\ref{tab:N}, we compare our column densities of N$_2$H$^+$ and N$_2$D$^+$ to those measured by \citet{Punanova2016} towards the dust emission peaks. The positions of their observations are marked with white crosses in our maps. Our column densities are lower then those in \citet{Punanova2016}, although not systematically and some agree within the uncertainties, with the only exception of N$_2$D$^+$ towards Oph-F where our column density is higher. This is likely due to the fact that our adopted $T_{\rm ex}$ or measured optical depths are lower than those measured in \citet{Punanova2016}. The reason why their $N_{\rm tot}$(N$_2$D$^+$) in Oph-F is lower is that they adopted $\tau=0.1$ for the N$_2$D$^+$ line while we measured $\tau$=1.6 towards that position. Our maps also show that some of the pointings in \citet{Punanova2016} missed the line emission peaks.
Our NH$_3$ column density maps agree with the map in \citet{Frisen2017}, which is expected since we used their data and the same line analysis method. 

Since some lines are subthermally excited, with the LTE approximation, we might underestimate the column densities \citep[e.g.,][]{Shirley2015}. The higher transitions have higher critical densities (see Table~\ref{tab:cof}) and also have more tendency to subthermal excitation, that is either way underestimated column density. The combination of a ground transition of hydrogenated isotopologue and higher transition of deuterated isotopologue would then lead to an underestimated deuterium fraction. However, the example of cores H-MM2 and I-MM2, where we have both (1--0) and (2--1) N$_2$D$^+$ transitions observed, shows the opposite result (see Table~\ref{tab:single-point}), where $N_{\rm tot}^{\rm N_2D^+(2-1)}>N_{\rm tot}^{\rm N_2D^+(1-0)}$ and agree within the errors. With RADEX \citep{vanderTak2007}, we estimated the column densities of the species towards the dust peaks of the cores, where volume density is measured, and found the difference within 10--20\%, both smaller and larger values. Non-LTE estimates of the column densities require the non-LTE radiative transfer modelling and thus the data on volume density distribution, that is the farther from the dust peak, the more model assumptions on the source geometry are involved. Thus, we consider LTE approximation the most optimal for column density mapping of dense cores. 

\subsection{Deuterium fraction}

\begin{figure*}
\centering
\includegraphics[scale=0.4]{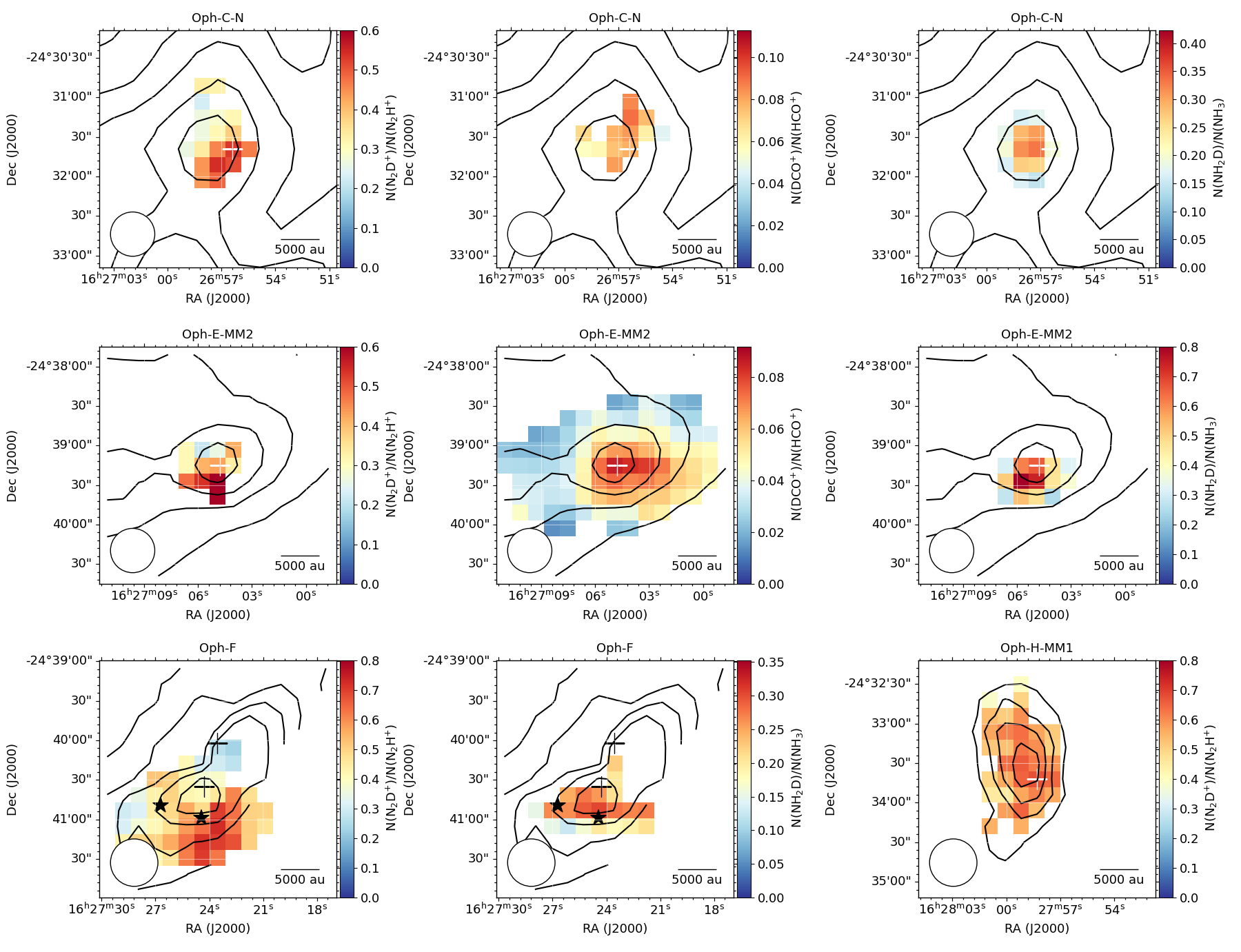}
\caption{The deuterium fraction maps towards Oph-C-N, Oph-E-MM2, Oph-F and Oph-H-MM1. The molecular hydrogen column density is shown by black contours. The first contour starts at 1.5$\times$10$^{22}$~cm$^{-2}$ with a contour step of 0.5$\times$10$^{22}$~cm$^{-2}$. {\it Top:} Oph-C-N; {\it centre:} Oph-E-MM2; {\it bottom:} Oph-F and Oph-H-MM1. The beam size is shown in the bottom left corner of each map. The crosses show the positions of the single pointing observations in \citet{Punanova2016}. The stars show the positions of the YSOs \citep[YLW15, Class~0$+$I, CRBR~2422.8-3423, Class~I;][]{Young1986,Kirk2017,Comeron1993,Bontemps2001} in Oph-F.}

\label{pic:Oph-C-N_R}
\end{figure*}

To estimate the deuterium fraction, we divided pixel by pixel the resulting column density maps of deuterated isotopologues by those of hydrogenated isotopologues. To calculate the column density of HCO$^+$, we applied the $^{12}$C/$^{13}$C=77 isotopic ratio for local ISM \citep{Wilson1994_HCO} to our $N$(H$^{13}$CO$^+$) maps and suggested that fractionation of $^{13}$C is negligible. We estimated the total deuterated ammonia column density based on our column density of para-NH$_2$D using o/p(NH$_2$D) = 3:1 obtained by \citet{Harju2017}. Figure~\ref{pic:Oph-C-N_R} shows the maps of the deuterium fractions in all tracers of the studied cold dense cores. Table \ref{tab:single-point} shows the deuterium fraction $R_D$(N$_2$D$^+$/N$_2$H$^+$) derived from the single-pointing observations towards Oph-H-MM2 and Oph-I-MM2.

Deuterium fractions appear high over entire maps and show a significant difference in the values of nitrogen-bearing $R_D$(N$_2$H$^+$) and $R_D$(NH$_3$) (peak value 0.6--0.8) and carbon-bearing $R_D$(HCO$^+$) (peak value 0.06--0.10) species. Despite the small extent, the obtained maps show that the deuterium fraction increases towards the dust peaks of the cores as it was observed in other dense cores \citep[e.g. in L1544, L1521F, L183;][]{Caselli2002,Crapsi2004,Crapsi2007,Pagani2007,Pagani2009b,Chacon-Tanarro+2019,Redaelli2019}, most prominently seen in the largest map of $R_D$(HCO$^+$) towards Oph-E-MM2. Only in Oph-F, where protostars are present, the maximum $R_D$(N$_2$D$^+$/N$_2$H$^+$) is away from the dust peak. However, $R_D$(N$_2$H$^+$) is still high ($\simeq0.4$) towards the protostars, and $R_D$(NH$_3$) is the highest towards one of the protostars. This means that the early stage protostars (Class~0--I) have not yet heated the dense gas of the core, or the heating of gas and dust has a very limited extent, diluted by our beam, or that they are not embedded in this core, being in the background. Towards the cores where the map size allows to see it (Oph-F and Oph-H-MM1), the deuterium fraction of N$_2$H$^+$ is high ($R_D>0.4$) over a large area (several beam sizes) which excludes the idea of very compact depletion zones with high deuteration proposed in \citet{Punanova2016}. 

We consider carbon-bearing H$^{13}$CO$^+$ and DCO$^+$ core shell tracers since they are depleted at high gas densities due to CO freeze-out, and nitrogen-bearing NH$_3$, NH$_2$D, N$_2$H$^+$, and N$_2$D$^+$ core centre tracers since they stay abundant at high densities of cold cores \citep[e.g.,][]{Caselli2002,Tafalla2006} and only get depleted at densities $\geq10^6$~cm$^{-3}$ \citep{Caselli2022}. However, the large-scale Green Bank Ammonia Survey showed that NH$_3$ traces also the low-density gas of the clouds \citep{Frisen2017}. We still consider NH$_3$ as a dense gas tracer since it stays undepleted within the cores (as indicated, e.g., by the lines, narrowing towards the core centre, see~Fig.~\ref{pic:Sigma_Pineda}) and recall that it also traces the low-density gas on the line of sight. Ammonia begins to deplete only at high densities ($>2\times10^5$~cm$^{-3}$) and smaller spatial scales \citep[$<1000$--3000~au][]{Pineda2022,Caselli2022}.

\begin{figure*}
\centering
\includegraphics[scale=0.4]{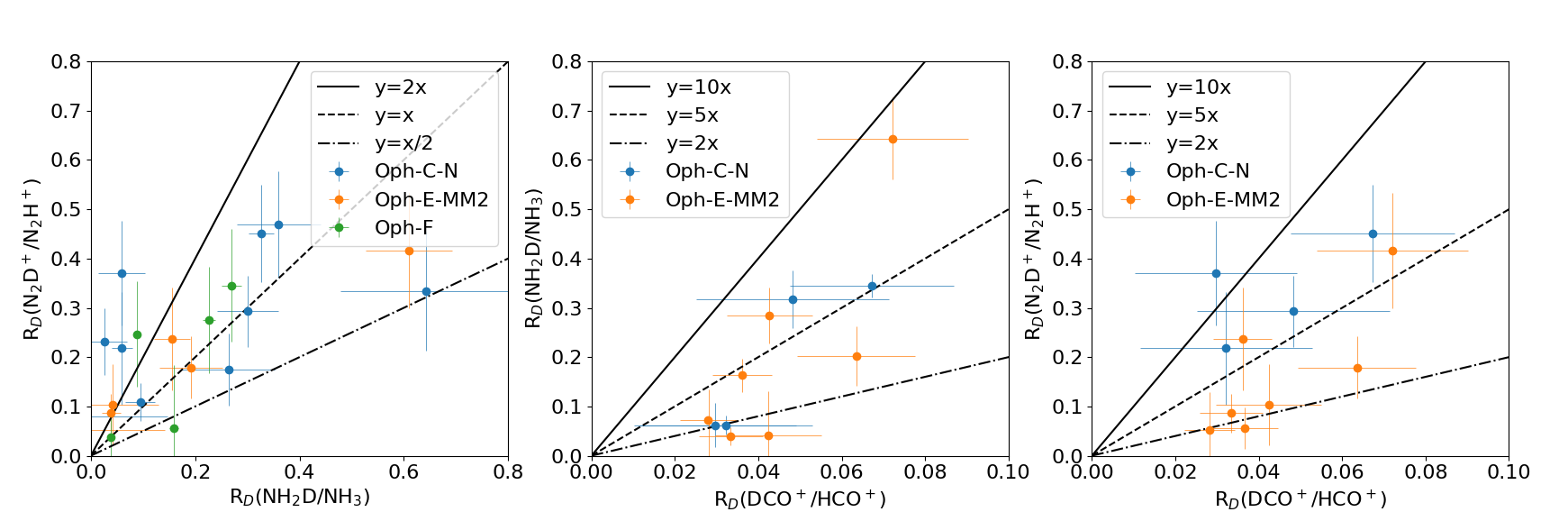}
\caption{The comparison of the deuterium fractions of the three species: {\it left:} $R_D$(N$_2$D$^+$/N$_2$H$^+$) versus $R_D$(NH$_2$D/NH$_3$); {\it centre:} $R_D$(NH$_2$D/NH$_3$) versus $R_D$(DCO$^+$/HCO$^+$); {\it right:} $R_D$(N$_2$D$^+$/N$_2$H$^+$) versus $R_D$(DCO$^+$/HCO$^+$). The lines show the ratio of the deuterium fractions.}
\label{pic:R_R}
\end{figure*}

In Fig.~\ref{pic:R_R} we assess the difference in the deuterium fractions of nitrogen-bearing and carbon-bearing species. The lines show the difference in deuterium fraction between N$_2$H$^+$ and NH$_3$ (factors of 0.5, 1, and 2, left panel) and between the N- and C-bearing species (factors of 2, 5, and 10, central and right panels). The deuterium fraction in nitrogen-bearing species is 2--10 times greater (the median difference is a factor of 4.2) than the deuterium fraction in carbon-bearing species. The deuterium fraction in ammonia is lower than that in N$_2$H$^+$ (the median difference is a factor of 1.5). The difference in deuterium fraction between the tracers indicates that it is greater in the core center than in the envelope.

\subsubsection{Deuterium fraction and the physical parameters of the cores}

To study how the physical conditions within the cores affect deuterium fraction, we analysed the correlations between the deuterium fraction of each species and the molecular hydrogen column density ($N_{\rm tot}$(H$_2$)), the dust temperature ($T_{\rm dust}$), the gas temperature ($T_{\rm gas}$) and gas turbulence, represented by the non-thermal component of the velocity dispersion ($\sigma_{\rm NT}$). Kinetic temperature rules the H$_2$ $o/p$ ratio and thus the deuterium chemistry. The correlation between deuterium fraction and the temperature is predicted by theoretical works \citep[e.g.,][]{Pagani2009b,Parise2011,Kong2015}. We used the dust temperature and molecular hydrogen column density maps from \citet{Ladjelate2020} based on the Herschel/SPIRE dust observations at 250~$\mu$m, 350~$\mu$m, and 500~$\mu$m, and the gas temperature map of L1688 from \citet{Frisen2017}. The non-thermal velocity dispersion was measured in our fits using the gas kinetic temperature from \citet{Frisen2017} as described in Section \ref{sec:spparam}. The non-thermal component is associated with turbulence, which increases the collisions of particles increasing the chances of species destruction and thus the number of deuterated molecules is decreased. The physical parameter maps were convolved to the common resolution (33.6$^{\prime\prime}$) of our data, for consistency. 

We fitted the linear approximations to estimate the correlation between the deuterium fractions and the physical parameters. We found no significant correlations between the deuterium fraction, the gas and dust temperature, molecular hydrogen column density ($N_{\rm tot}$(H$_2$)) and the non-thermal component. The gas temperature $T_{\rm gas}$ dynamical range of only 2~K is probably too small to show any correlation. The lines are mainly subthermal, and the maps of deuterium fraction do not reach the `transition to coherence' \citep{Pineda2010,Chen2019}, which means also small variation in turbulence. The correlations between deuterium fraction, molecular hydrogen column density ($N_{\rm tot}$(H$_2$)) and dust temperature have large uncertainties in the parameters of approximating linear functions due to the small number of points. Maps of deuterium fraction spanning beyond the `transition to coherence' are needed to show the correlation between deuterium fraction and physical parameters.

\subsection{CO depletion}\label{CO}
Deuterium fraction is expected to increase with CO freeze out \citep{Bacmann2003,Caselli2008,Kong2015}, since CO, the second most abundant molecule in molecular cloud gas after H$_2$, reacts and destroys the H$_3^+$ ion, which is needed to support the deuterium chemistry via the H$_3^++$HD $\rightleftharpoons$ H$_2$D$^++$H$_2$ reaction. To test this prediction with our data set, we took the $^{13}$CO(1--0) observations from the COMPLETE survey \citet{Ridge2006} and produced the CO column density maps. The spectral lines of $^{13}$CO(1--0) were partly optically thick and had several unresolved velocity components. We were unable to estimate the optical depth of the line, thus we calculated column density as if the lines were optically thin, as described in Sect.~\ref{sec:colden} and since the lines are partly optically thick the column density therefore represents its lower limit. We converted the column density of $^{13}$CO to the column density of CO using the $^{12}$C/$^{13}$C fractional abundance of 77 \citep{Wilson1994_HCO}. We compared the obtained values of the column density towards the cores dust peaks to those measured via the C$^{17}$O observations from \citet{Punanova2016}. Our estimates of the CO column density done with $^{13}$CO(1--0) are 3--8 times lower than the values from \citet{Punanova2016}, depending on the core. Even with the upper limit, the distribution of $f_d$ is consistent with theory, it is higher towards the dust peak where the deuterium fraction should be the highest. However we note that while $^{13}$CO(1--0) is obviously optically thick towards the core dust peaks, its intensity decreases at lower densities, thus the estimated column density and $f_d$ are more likely to be accurate at the edges of the core maps. 

We took the reference CO abundance ($X$(CO)$_{\rm ref}$ = 2.69$\times$10$^{-4}$) for Ophiuchus from \citet{Lacy1994} and plotted the maps of the upper limit of the CO depletion:
\begin{equation}
f_d({\rm CO}) =\frac{X({\rm CO})_{\rm ref}}{X({\rm CO})_{\rm obs}},
\end{equation}
where $X$(CO)$_{\rm obs}$=$N$(CO)/$N$(H$_2$). Figure~\ref{pic:COfd} shows the maps of the CO depletion upper limit for our cores based on $^{13}$CO. The maximum values of CO depletion are given in Table~\ref{tab:fdCO}. Since we have only lower limits on the CO column density, the distribution of the CO depletion is rather affected by the column density of H$_2$, and the correlation between the deuterium fraction and CO depletion factor is very similar to that between the deuterium fraction and molecular hydrogen column density. The analysis of C$^{17}$O(1--0) presented in \citet{Punanova2016} suggested that there is no CO depletion towards the dust peaks of Oph-C-N, Oph-E-MM2, and Oph-F. Despite the upper limits of $f_d$ towards the dense cores, the advantage of these maps is that they show moderate depletion ($f_d$=1--2) towards the cores outskirts where the $^{13}$CO(1--0) line have low optical depth and where the column density of CO might be measured accurately.

\begin{table}
\begin{center}
\begin{tabular}{| l | c |}
Core & $f_{d{\rm max}}$(CO) \\ \hline
Oph-C-N & $<$7.70$\pm$0.05 \\ 
Oph-E-MM2 & $<$6.96$\pm$0.06 \\ 
Oph-H-MM1 & $<$8.41$\pm$0.07\\ 
Oph-F & $<$9.42$\pm$0.07\\ 
\hline
\end{tabular}
\end{center}
\caption{The upper limits of the CO (from $^{13}$CO) depletion towards the observed starless cores.}\label{tab:fdCO}
\end{table}

\subsection{Ionisation fraction}
The ionization degree indicates the strength of the coupling between the magnetic field and the dense core. A strong magnetic field can affect the stability of the dense core, thereby preventing contraction or fragmentation. The simplest way to estimate the ionization degree is to count all available observed ions:
\begin{equation}\label{eqi}
    x(e) = \frac{N(\mathrm{HCO^+})+N(\mathrm{DCO^+})+N({\mathrm{N_2H^+})+N(\mathrm{N_2D^+})}}{N(\mathrm{H_2})},
\end{equation}
as it is described in \citet{Caselli2002}. HCO$^+$ and N$_2$H$^+$, and their deuterated isotopologues are the most abundant ions in dense gas after H$_3^+$, the latter we did not observe, so this method can only estimate the lower limit of the ionization degree. HCO$^+$ contributes the most to our estimate of the ionisation degree. In our data set, we observed all four ions towards two cores, Oph-E-MM2 and Oph-C-N. However, for Oph-C-N, HCO$^+$, the most abundant of the four ions, is represented by the H$^{13}$CO$^+$(2--1) line at 173.5~GHz, where atmospheric absorption is strong, thus the noise temperature is high, and the high uncertainty in the integrated intensity propagates to the column density and the estimate of the ionization degree, so the uncertainty of the latter is very high. Thus we performed the ionization analysis only for Oph-E-MM2. The left panel of Fig.~\ref{pic:ion} shows the ionization degree in Oph-E-MM2. The maximum ionization degree ($x(e) = 3.05\pm0.9\times 10^{-9}$) is similar to that of the prototypical pre-stellar core L1544 \citep[$x(e)\sim 10^{-9}$;][]{Caselli2002}.

We also estimated the ionization fraction and the cosmic ray ionization rate for Oph-E-MM2 using the analytical definition of [DCO$^+$]/[HCO$^+$]$\equiv R_D$ and [HCO$^+$]/[CO]$\equiv R_H$ based on a simple steady-state model, where $R_D$ and $R_H$ are defined via formation, destruction and dissociation reaction rates, ionization rate, ionization degree and HD abundance \citep[for the details, see Sect.~3 in][]{Caselli1998}. Following this approach, we estimate the ionization degree at the temperature of 10~K, characteristic for a cold core and already implemented in the model:
\begin{equation}\label{eq6}
    x(e) = \frac{2.7\times10^{-8}}{R_D} - \frac{1.2\times10^{-6}}{f_d},
\end{equation}
and cosmic rays ionization rate $\zeta$:
\begin{equation}\label{eqz}
    \zeta = \left[7.5\times10^{-4}x(e)+\frac{4.6\times10^{-10}}{f_d}\right]x(e)\times n(\mathrm{H_2})\times R_H,
\end{equation}
where $f_d$ is the depletion factor, the numerical coefficients $2.7\times10^{-8}$ and $1.2\times10^{-6}$, $7.5\times10^{-4}$ and $4.6\times10^{-10}$ are obtained by substituting reaction constants following \citet{Caselli1998}.
The value of volume density $n$(H$_2$)=4$\times$10$^5$~cm$^{-3}$ for Oph-E-MM2 was taken from \citet{Ward-Thompson1999}; for $N$(CO) we used $N$(C$^{17}$O) from \citet{Punanova2016} and the isotopic ratios $^{16}$O/$^{18}$O=560 \citep{Wilson1994_HCO} and $^{18}$O/$^{17}$O=4.11 \citep{Wouterloot2005}.

Figure~\ref{pic:ion} shows the ionization degree (central panels) and ionization rate (right panels) towards Oph-E-MM2 (the upper panels show the values, the lower panels show the uncertainties). The minimum values for the ionization degree are (2.0$\pm$0.2)$\times$10$^{-8}$ for the steady-state model from \citet{Caselli1998} and the lower limit of (1.00$\pm$0.03)$\times$10$^{-9}$ for observations. The ionization rate is the lowest close to the core centre: (2$\pm$8)$\times$10$^{-17}$~s$^{-1}$. However, towards the central pixels, $x(e)$ and thus $\zeta$ become negative, since the $R_D$ increases and the $f_d$ remain constant (we use one value). In principle, the dust peak is the only position where we could correctly estimate $x(e)$ and $\zeta$ since it is the only position with reliable $f_d$ measurement. The value is so small ($f_d$=1.25) that it leads to a negative result in Eq.~\ref{eq6}. To obtain here a positive $x(e)$, $f_d$ needs to be at least 3.83. This is another indication of the fact that the total undepleted abundance of CO in the Ophiuchus molecular cloud may be larger than a `canonical' value, common for the local ISM \citep[between 1 and 2$\times 10^{-4}$, e.g.,][]{Frerking1982,Lacy1994} and possible variation of metallicity from one cloud to another. 
However, towards the core outskirts the value $f_d$=1.25 might be adequate, and the estimates of $x(e)$ and $\zeta$ may be correct. Unlike the first method that gives the lower limit of the ionization degree (3.8$\times 10^{-9}$), the model estimates the total value, which is two-three orders of magnitude higher, $\sim 10^{-6}$ (similar to the one, $\sim 10^{-6.5}$, recently found towards NGC~1333 by Pineda et al., submt.), and shows that the most abundant ions are not counted, possibly H$_3^+$ and its isotopologues. Although such high ionization degree ($\sim 10^{-6}$) is rather characteristic for a surrounding translucent cloud, not dense cores \citep[e.g.][]{Bron2021}, there’s some evidence that L1688 has a higher local ISRF than, e.g., towards L1544, given its nearby YSO population and warmer gas temperatures overall.
To apply correctly the simple model from \citet{Caselli1998,Caselli2002} one needs to correctly estimate CO depletion that remains challenging for L1688 and requires mapping of the rare CO isotopologues down to low densities where CO is undepleted. 

 \subsection{Chemical modelling} \label{sec:model}
We use the gas-grain chemical model described in \citet{Sipila2015_model} to reproduce the observed column densities and deuterium fractions of the observed species. The model distinguishes between spin states of hydrogen and deuterium bearing species, with branching ratios derived by applying selection rules under the principle of the conservation of nuclear spin. 
Ortho-hydrogen affects the deuterium fractionation because the deuteron-proton exchange reaction with it is reversible even at low temperatures \citep{Pagani1992}. The model considers two phases, gas phase and dust surface, and connects the two via adsorption and several mechanisms of desorption. Chemical reactions on the dust surfaces are also included.

The elemental abundances are taken from \citet{Sipila2015_model}, and correspond to the low-metal composition \citep[e.g.,][]{Wakelam2008}, which is more consistent with molecular cloud evolution and the low electron fractions observed there \citep{Graedel1982}. \citet{Harju2017} found that they can not reproduce the observed ammonia abundances with the same chemical model and suggested using a higher nitrogen abundance (N/H=5.3$\times10^{-5}$ instead of 2.14$\times10^{-5}$), that improved the results but did not produce a perfect match. We explored how the initial nitrogen abundance affects the resulting abundances of the studied species and, in addition to the low-metal N/H=2.14$\times10^{-5}$, used even higher than that in \citet{Harju2017} nitrogen abundance N/H=7.60$\times10^{-5}$ that corresponds to the high-metal elemental abundances \citep{Cardelli1993,Wakelam2008}, to find the best match with the observations. 

\subsubsection{Physical model}

We model the dense cores as static spheres with gas and dust temperature and density profiles. The dust emission peaks were chosen to be the core centres. We used column density $N$(H$_2$) and dust temperature $T_{\rm dust}$, based on the Herschel observations, from \citet{Ladjelate2020}. To obtain discrete profiles of $N$(H$_2$) and $T_{\rm dust}$, we averaged their values within concentric rings with a width of 1285~au which is equal to the pixel size \citep[12$^{\prime\prime}$ at the distance of 119~pc;][]{Zucker2019} in our maps. We used the method described, e.g., in equation~1 of \citet{Hasenberger2020} to obtain the gas density, $n$(H$_2$), from the molecular hydrogen column density. We convert $N$(H$_2$) to interstellar extinction, $A_V$, applying a factor of 9.4$\times10^{20}$ \citep{Frerking1982}. For the density towards the centre, we took the volume densities obtained by \citet{Pattle2015}, since they are based on the combination of the Herschel (250~$\mu$m, 350~$\mu$m, 500~$\mu$m) and SCUBA2 (450~$\mu$m and 850~$\mu$m) observations, that is on dust emission at longer wavelength that better probe cold dust in core centre. We added one extra point at a distance of 642~au from the centre to have the central 1285~au-wide circle. This gives 9 points from the center to the edge of the core with a radius of 9000~au, so the model cores would cover entirely the extent of our maps. The dust temperature slightly decreases towards the core centers, by 2--5~K compared to the core edge (where $T_k$=15.5--17.5~K). The molecular hydrogen volume density increases towards the core centers by an order of magnitude. 

\begin{figure*}
\centering
\includegraphics[scale=0.38]{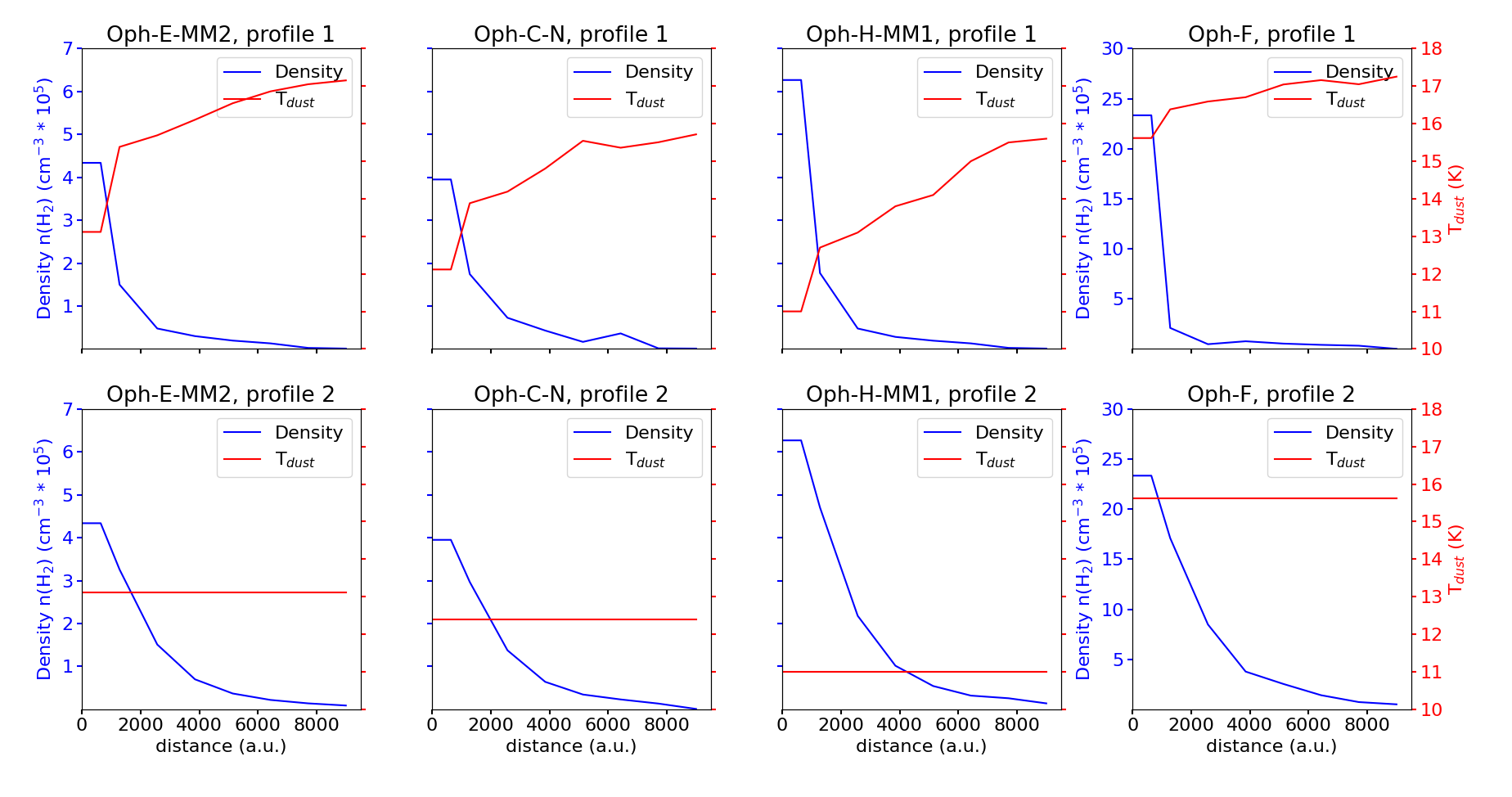}
\caption{ Model profiles of gas density and dust temperature: profile~1 is based on the $N$(H$_2$) and $T_{\rm dust}$ from \citet{Ladjelate2020} with a central density and temperature from \citet{Pattle2015}; profile~2 is based on the analytical profile from \citet{Tafalla2002} and $n$(H$_2$) from \citet{Pattle2015}, $T_{\rm dust}$ is constant, from \citet{Pattle2015}.} %
\label{pic: model_prof}
\end{figure*}

We also used an alternative analytical profile based on the central density from \citet{Pattle2015} and theoretical radial density distribution from \citet{Tafalla2002}. The dust temperature is constant in the alternative model. Figure~\ref{pic: model_prof} shows the gas density and the dust temperature profiles for our cores. Profiles~1 are based on the data from \citet{Ladjelate2020} and profiles~2 are based on the method from \citet{Tafalla2002}.

\subsubsection{Conversion of abundances to column densities}\label{sec:conversion}
The chemical model produces the abundances of species in each position independently. To compare the model results to our observed column densities, we convert the modelled abundances to the line of sight column densities, considering the spherical geometry and convolution with the telescope beam. The total modelled column density on the line of sight is calculated as the sum of the column densities multiplied by the layer thickness in point $i$ on the line of sight (the profile has $z=8$ points): 
\begin{equation}
N_{\rm tot} = \sum\limits_{i=1}^z{(X_{i}\times n_{i}({\rm H}) + X_{i-1}\times n_{i-1}({\rm H})) \Delta r_i},
\label{form:1}
\end{equation}
where $X_{i}$ is the abundance in point $i$, $n_{i}$(H) is the concentration of atomic hydrogen in point $i$, $\Delta r_i$ is the distance between points $i$ and $i$--1. For the beam towards the centre:
\begin{multline}
N_{\rm tot}^0 = X_{0}\times n_{0}({\rm H}) \times \Delta r + \\
+\sum\limits_{i=1}^z{(X_{i}\times n_{i}({\rm H}) + X_{i-1}\times n_{i-1}({\rm H}))  \Delta r},
\label{form:2}
\end{multline}
where $X_{0}$ is the abundance in the central point, $n_{0}$(H) is the concentration of atomic hydrogen in the central point, and $\Delta r$ is the 1285~au step of our density profile which is here the distance between any two neighboring points on the line of sight towards the centre.

\subsubsection{Convolution with a beam}

For a correct comparison with observational results, we convolved the modelled column densities with the beam size of our observations, 33.6$^{\prime\prime}$.
We have constructed Gaussian distribution normalized by one for each point with FWHM of 33.6$^{\prime\prime}$. The smoothed column density for each point along the profile is then: 
\begin{equation}
N_{\rm tot\_sm} = \sum\limits_{j=0}^z{N^j_{\rm tot} \times I^j_{\rm beam}},
\end{equation}
where $N^j_{\rm tot}$ -- column density in point~$j$, $z=17$ -- points from $-8$ through 0 to 8 is needed for the sum over the entire length of the Gaussian, $I^j_{\rm beam}$ -- the intensity of the Gaussian in point~$j$. 
As a result, we have nine points of the column density convolved with our beam in the plane of the sky along the profile from the center to the edge of the core, for each species.

\subsubsection{Model results: temporal evolution and chemical age}
We ended up with four models with different input parameters: profile~1 with $X$(N)=7.60$\times$10$^{-5}$ and $X$(N)=2.14$\times$10$^{-5}$ and profile~2 with $X$(N)=7.60$\times$10$^{-5}$ and $X$(N)=2.14$\times$10$^{-5}$.
Figures~\ref{pic: model_cn}, \ref{pic: model_emm2}, \ref{pic: model_f}, and~\ref{pic: model_hmm1} compare the modelled and observed column densities and deuterium fractions for all cores. The first and second (upper and middle upper) rows of panels show how the column densities and deuterium fractions change with the evolution time. The third and fourth (middle lower and lower) rows of panels show how the column densities and deuterium fractions change with the distance from the core centres at the chosen age.

In all four variants (physical profiles and initial abundance of N), the model showed an increase with time in the column densities of N$_2$H$^+$ and of the deuterated species -- N$_2$D$^+$, DCO$^+$, and NH$_2$D. N$_2$D$^+$ showed the steepest increase with no decreasing at the early ages ($<10^5$~yr). Both NH$_3$ and NH$_2$D showed a plateau or even a decrease around a few $10^3$--$10^4$~yr. The peak of gas-phase NH$_3$ column density is around $\sim10^3$~yr. The HCO$^+$ column density slightly increases over the time.

To compare the model results with observations, we need to set the age of the cores. The CO depletion factor allows to estimate the core chemical age, since CO chemistry is fairly simple and well established \citep[as was done in, e.g.,][]{Vasyunin2017}. However, we have only upper limits of the CO depletion factor, so we did not use it to estimate the age. We still plot the modelled CO depletion factor (in the second rows of Fig.~\ref{pic: model_cn}, \ref{pic: model_emm2}, \ref{pic: model_f}, and~\ref{pic: model_hmm1}) over the chemical evolution to discuss later its correlation with deuterium fraction. To estimate the model CO depletion, we need to know the undepleted abundance of CO. For that, we run the chemical model with $T_{\rm gas}$=20~K, $A_V$=2, $n$(H$_2$)=200~cm$^{-3}$ until 10$^6$~yr, and use the maximum gas-phase CO abundance reached in the diffuse cloud, $X({\rm CO})_{\rm cloud}=2.5\times 10^{-5}$. We estimate the model CO depletion factor as
\begin{equation}
f_d({\rm CO}) =\frac{X({\rm CO})_{\rm cloud}}{N({\rm CO})_{\rm gas}/(2N({\rm H_2})+N({\rm H}))},
\end{equation}
where $N({\rm CO})_{\rm gas}$ is the simulated CO column density of CO in the gas phase, $N({\rm H_2})$ and $N({\rm H)}$ are the column densities of molecular and atomic hydrogen, all produced by the chemical model of the core. 

We then suggested that the observed high the deuterium fractions might be the highest possible in our cores; and that deuterium fraction of N$_2$H$^+$ is the most sensitive to the change of the physical conditions. Thus, to compare the modelled and observed column density profiles, we chose the ages of the first $R_D$(N$_2$D$^+$/N$_2$H$^+$) deuterium fraction peaks. 
Based on the peaks of the deuterium fraction $R_D$(N$_2$D$^+$/N$_2$H$^+$), we obtained the chemical ages of cold dense cores. The ages vary from one the model to the other (see Fig.~\ref{pic: model_cn}, \ref{pic: model_emm2}, \ref{pic: model_f}, \ref{pic: model_hmm1}), but not significantly. The median ages are 2.65$\times10^{5}$~yr, 2.65$\times10^{5}$~yr, 0.82$\times10^{5}$~yr, and 1.80$\times10^{5}$~yr for Oph-C-N, Oph-E-MM2, Oph-F and Oph-H-MM1, respectively. 
The deuterium fractions of N$_2$H$^+$, NH$_3$, and HCO$^+$ show peaks at different chemical ages, all of them are around (0.8--4.1)$\times$10$^{5}$~yr. The age of $\sim$10$^{5}$~yr is considered to be a typical chemical age for cold cores \citep[e.g.,][]{Caselli1998,Vasyunin2017,Harju2017}. We nevertheless analyse each variant of the model (with two nitrogen abundances and two physical profiles) and plot the radial distribution of the column densities at the ages found within each model. The ages are shown on top of each column in Fig.~\ref{pic: model_cn}, \ref{pic: model_emm2}, \ref{pic: model_f}, and~\ref{pic: model_hmm1}. Chemical ages are not related to dynamical ages because our model has a static physical structure.

\subsubsection{Model results: column densities and deuterium fractions}

While all the modelled column densities agree with the observed ones within one order of magnitude, which falls into the model accuracy \citep{Sipila2015_model}, we still see the following trends in the model results. We found that modelled column densities of HCO$^+$ and DCO$^+$ agree with the observed ones for both physical profiles and both nitrogen elemental abundances, with an underestimate for HCO$^+$ and an overestimate for DCO$^+$ at the edges of the cores (outer 6000--9000~au, see Fig.~\ref{pic: model_cn}, \ref{pic: model_emm2}, \ref{pic: model_f}, and~\ref{pic: model_hmm1}). 
Regardless of the density and temperature profiles applied, with the low nitrogen abundance, the modelled column densities of N$_2$H$^+$ and N$_2$D$^+$ are underestimated compared to the observations. The high nitrogen abundance gives an overestimate in the N$_2$H$^+$ column density and either an overestimate or good agreement in the N$_2$D$^+$ column density. 
The modelled column density of NH$_3$ is overestimated by all variants of the model except for the models with the low nitrogen abundance for Oph-F, where the modelled column densities agree with the observed ones. The modelled column densities of NH$_2$D agree with the observed ones for the cores Oph-E-MM2 and Oph-C-N with small over- and underestimates, with all four the models. All four models underestimate the column density of NH$_2$D for Oph-F, models with the high nitrogen abundance give a better agreement with observations.  

Although the model deuterium fractions at the chosen chemical ages agree with the observed deuterium fractions within the uncertainty of the model, we were able to identify the most suitable models.
The modelled deuterium fraction $R_D$(N$_2$D$^+$/N$_2$H$^+$) either agrees or underestimates the observed one for all cores, but profile~2 gives a better result for the entire core while the Herschel-based profile~1 shows agreement only towards the core centres. The modelled deuterium fractions $R_D$(N$_2$D$^+$/N$_2$H$^+$) show a similar distribution along the core radii as the observed deuterium fractions.

The modelled deuterium fractions R$_D$(NH$_2$D/NH$_3$) are much smaller (by a factor of 8--17) than the observed ones. The modelled deuterium fractions $R_D$(DCO$^+$/HCO$^+$) either overestimate or agree (towards Oph-E-MM2 with profile~1, both $X$(N), and towards Oph-C-N with $X$(N)=7.60$\times$10$^{-5}$, both profiles) with the observed ones. 

The modelled CO depletion factor reaches 4--7 at the chosen chemical ages. Figure~\ref{pic:COfd} shows the upper limit on the CO depletion factor (described in Sect.~\ref{CO}) and these values are only slightly higher ($f_d\leq$8--10) than those in the model. 

The analytical profile~2 causes an increased abundance of deuterated species. The high nitrogen abundance makes the biggest difference in the modelled column densities: all nitrogen-bearing species show higher column densities, the carbon-bearing species show lower column densities or remain unchanged. In cores Oph-F and Oph-H-MM1 the initial nitrogen abundance affects only the column densities of the N-bearing species, the column densities of the C-bearing species are almost the same in the four variants of the model. The results in these two cores do not depend on the chosen physical profile.


\section{Discussion}\label{sec:discussion}
\subsection{Our line fitting approach}

The problem with fitting hfs structure to the spectra with moderate signal-to-noise ratio is the large uncertainty of the optical depth. One approach to treat such spectra of lines with low optical depths is to consider the lines optically thin \citep[e.g.,][]{Crapsi2005}. However, a reasonable value for optical depth is important to adequately estimate column densities. We used the Monte-Carlo method to explore the mutual correlation between $T_{\rm ex}$ and $\tau$, found the most probable $T_{\rm ex}$ for each transition in each core, and applied a $T_{\rm ex}$-constrained fit to estimate $\tau$. An incorrect $T_{\rm ex}$ also affects the estimate of column density (see Sect.~\ref{sect:line_analysis}). As a result, towards the dust peaks, we obtained lower column densities of both N$_2$H$^+$ and N$_2$D$^+$ than those in \citet{Punanova2016} since either $\tau$ or $T_{\rm ex}$ in our fits were lower (see Sect.~\ref{sec:colden}). 


We compared our results of the deuterium fraction $R_D$(N$_2$H$^+$) to the values from \citet{Punanova2016} (see Table~\ref{tab:N}). Despite the fact that we got lower values for the column densities, the deuterium fractions towards Oph-C-N, Oph-E-MM2, and Oph-H-MM1 agreed within the uncertainties (our values are higher) while towards Oph-F our value was higher since we measured the optical depth of the N$_2$D$^+$(2--1) line, and \citet{Punanova2016} assumed the line to be optically thin, that lead to an underestimate of $N_{\rm tot}$(N$_2$D$^+$) and the deuterium fraction. 

The map of $R_D$(N$_2$H$^+$) towards Oph-H-MM1 was presented in \citet{Petrashkevich2020}, and they also assumed the N$_2$D$^+$(1--0) emission was optically thin. Our results show the optical depth of the N$_2$D$^+$(1--0) line $\tau\leq5$. It means that \citet{Petrashkevich2020} underestimated the column density of N$_2$D$^+$ due to the assumption of the optically thin lines. With our fitting method, we obtained a factor of $\sim$2 higher column densities of N$_2$D$^+$, 16\% higher column densities in N$_2$H$^+$, and as a result, a factor of 1.7 higher deuterium fraction. 

To conclude, a free-parameter fit gives a better result but requires high quality spectra. Our approach is a reasonable compromise to fit noisy spectra with moderate or low signal-to-noise ratio, which are always present in mapping, and results in more realistic estimates of column densities than a mere assumption of optically thin emission.

\begin{table}
\begin{center}
\begin{tabular}{lccc}\hline
Core 	& $R^{\rm max}_D$         & $R^{\rm max}_D$  & $R^{\rm max}_D$  \\
        & (N$_2$D$^+$/N$_2$H$^+$) & (NH$_2$D/NH$_3$) & (DCO$^+$/HCO$^+$) \\ \hline
Oph-C-N & 0.53$\pm$0.14 	  & 0.32$\pm$0.06    & 0.090$\pm$0.017  	\\	
Oph-E-MM2 & 0.62$\pm$0.19  	  & 0.81$\pm$0.16    & 0.086$\pm$0.018  	\\	
Oph-F 	& 0.73$\pm$0.18 	  & 0.30$\pm$0.05    & --			\\	
Oph-H-MM1 & 0.67$\pm$0.17 	    & --	     & --				\\	
Oph-H-MM2 &0.38$\pm$0.11 	    & --	     & --				\\	
Oph-I-MM2 & 0.26$\pm$0.09       & --	     & --				\\
\hline
\end{tabular}
\end{center}
\caption{Maximum values of the deuterium fraction towards the observed starless cores.} \label{tab:Rmax}
\end{table}

\subsection{High deuterium fraction}
We obtained remarkably high deuterium fractions in N$_2$H$^+$ and NH$_3$ towards the studied cores, 0.26--0.81 (see Table~\ref{tab:Rmax}). Even considering an overestimate caused by our line fitting method, the deuterium fraction measured earlier towards these cores is high \citep[0.13--0.43, as measured via N$_2$D$^+$/N$_2$H$^+$, NH$_2$D/NH$_3$;][]{Punanova2016,Harju2017,Petrashkevich2020}. Moreover, the high $R_D$(N$_2$D$^+$/N$_2$H$^+$)$\geq0.4$ is observed over the area of several beam sizes towards Oph-F and Oph-H-MM1. Since the higher transitions of deuterated species have higher critical densities (see Table~\ref{tab:cof}) and likely trace columns of material shorter than the H-bearing ones, the deuterium fractions are underestimated. High deuterium fractions ($>$0.3) have been observed towards few cores: $R_D$(N$_2$D$^+$/N$_2$H$^+$)=0.70$\pm$0.12 in L183 in Serpens \citep{Pagani2007}, $R_D$(NH$_2$D/NH$_3$)=0.5$\pm$0.2 in L1544 in Taurus \citep{Crapsi2007}, $R_D$(N$_2$D$^+$/N$_2$H$^+$)=0.44$\pm$0.0.08 in Oph~D in Ophiuchus, also L1688 \citep{Crapsi2005}; $R_D$(NH$_2$D/NH$_3$)=0.35--0.50 in C16 \citep{Galloway-Sprietsma2022}. Towards the majority of cold dense cores the deuterium fraction is $R_D$(N$_2$D$^+$/N$_2$H$^+$)=0.03--0.28 \citep[in Taurus, Aquila, Perseus, Ophiuchus;][]{Crapsi2005,Emprechtinger2009,Friesen2013} and $R_D$(NH$_2$D/NH$_3$)=0.003--0.270 \citep[although most of the cores were observed with a larger 70$^{\prime\prime}$ beam;][]{Shah2001,Roueff2005,Galloway-Sprietsma2022}.

At the same time, in L1688, there are clumps with low or moderate deuterium fractions: Oph-A with $R_D<0.1$ and Oph-B2 with $R_D<0.2$ \citep{Friesen2010_3,Punanova2016}; these clumps contain many protostars \citep{DiFrancesco2008}, which explains the lower values. Protostars slow deuterium fractionation and start the reverse process by heating gas and dust. \cite{Friesen2010_3} presented the map of deuterium fraction, $R_D$(N$_2$D$^+$/N$_2$H$^+$), towards Oph-B2, a clump in L1688. They obtained the maximum deuterium fraction of 0.16. Several protostars embedded in Oph-B2, some of them associated with outflows \citep{Kamazaki2003}, could increase the temperature and ionization degree within the clump and cause the decrease in deuterium fraction \citep{Friesen2010_3}. Unlike Oph-B2, Oph-F and Oph-I-MM2, also containing embedded protostellar cores \citep{Kirk2017}, show high deuterium fractions, up to 0.3 and 0.7 in NH$_3$ and N$_2$H$^+$, respectively. Either the protostars are too young and have not yet affected the deuterium chemistry in the cores (or have affected small areas diluted by our 30$^{\prime\prime}\simeq4000$~au beam), or they are not embedded but projected onto the cores. 

The high deuterium fraction observed towards the cores supports the previously proposed idea that the cores in L1688 are close to the pre-stellar stage \citep{Punanova2016} and consistent with the age of several free-fall times. This might be a result of the dominating pressure of the cloud where cores are fully or mostly pressure-confined \citep{Pattle2015,Chen2019}. The gas in the cores evolve chemically while the mass is not enough to start the contraction \citep[the cores are gravitationally unbound;][]{Pattle2015,Kerr2019}, the relatively strong magnetic field \citep[$B_{\rm tot}$=75--135~$\mu$G in L1688;][]{Hu2023} also resists the contraction, so deuteration increases. The low CO depletion traced by C$^{17}$O observed towards the core dust peak contradicts this idea as well as the fact of high deuterium fraction observed across the cores. Possibly, our estimate of CO depletion in L1688 is incorrect, and one should apply a larger reference CO abundance (we apply the largest published in literature). A recent work of \citet{Punanova2022} also showed that a larger reference CO abundance is needed to correctly estimate CO depletion in dense cores. Then the picture will be consistent: CO is depleted, deuteration is enhanced, that is the cores exist in this quasistatic phase for time much longer than free-fall time.

\subsection{The three different gas density tracers}

We studied the tracers of different gas density -- carbon-bearing HCO$^+$ and nitrogen-bearing NH$_3$ and N$_2$H$^+$, and their deuterated isotopologues.

Carbon-bearing species form relatively fast and are abundant in the molecular cloud. The formation of nitrogen-bearing NH$_3$ and N$_2$H$^+$ occurs in denser regions at later stages of molecular cloud evolution, because N$_2$ formation requires longer time \citep[e.g.,][]{Hily-Blant2010}. Figure~\ref{pic:R_R} shows the difference between the three density tracers. Deuterium fraction of HCO$^+$ is smaller than that of NH$_3$ and N$_2$H$^+$ by a factor of 2--12. This is due to the higher abundance of CO in the gas phase and higher temperature in the core envelope. 

The similar difference in the deuterium fraction was observed towards the prototypical pre-stellar core L1544 for carbon-bearing \citep[$R_D$(DCO$^+$/HCO$^+$)=0.06$\pm$0.02; $R_D$(CH$_2$DOH/CH$_3$OH)=0.08 and $R_D$(D$_2$CO/H$_2$CO)=0.04$\pm$0.03;][]{Caselli2002,Redaelli2019,Chacon-Tanarro+2019} and nitrogen-bearing \citep[$R_D$(N$_2$D$^+$/N$_2$H$^+$)=0.24$\pm$0.02; $R_D$(NH$_2$D/NH$_3$)=0.5$\pm$0.2;][]{Caselli2002,Crapsi2007,Redaelli2019} species as in L1688. Also $R_D$(CH$_2$DOH/CH$_3$OH)=0.04--0.23 in the cores of B10/L1495 \citep{Ambrose2021}.

NH$_3$ is observed in molecular clouds at lower densities than N$_2$H$^+$ since it is excited and probably formed at lower densities
than N$_2$H$^+$. Because of this, we trace more material with NH$_3$ than that with N$_2$H$^+$ on the line of sight. At the same time, NH$_2$D and N$_2$D$^+$ are present and excited in almost the same compact and dense area, so using NH$_3$ to measure $R_D$, we divide by a larger column due to the excitation conditions (see Fig.~\ref{Contr}).
It is also possible that at high densities a decreasing ionization rate affects the difference in the deuteration of N$_2$H$^+$ and NH$_3$: N$_2$D$^+$ dissociates in a reaction with a free electron, while NH$_2$D needs a free electron to be formed out of the NH$_3$D$^+$ ion \citep[this reaction accounts for over 80\% of the formed NH$_2$D, according to the model;][]{Sipila2015_model}. 

The deuterium fraction map of Oph-F is not significantly affected by the embedded protostar and $R_D$(N$_2$D$^+$/N$_2$H$^+$) towards the protostars remain high ($\simeq0.4$). $R_D$(NH$_2$D/NH$_3$) remains unaffected. While we can assume that reverse deuterium fractionation \citep[e.g.,][]{Friesen2010_3} has started in N$_2$H$^+$, it has not started in NH$_3$. This also can be caused by locally increased ionization degree (due to the presence of a protostar) that affects N$_2$D$^+$ faster than NH$_2$D.

\subsection{Performance of the chemical model}
We model the chemical composition of the studied cold dense cores using the gas-grain chemical model pyRate3 presented in \citet{Sipila2015_model,Sipila2015_spin_state}. The model distinguishes the hydrogen isotopes, protium and deuterium, and takes into account hydrogen and deuterium spin states and thus the energy difference between the ortho- and para- states in the reactions containing up to five atoms of hydrogen or deuterium based on the spin selection rules. To reproduce the physical conditions, we used a model with two different molecular hydrogen density profiles and gas and dust temperatures profiles. The first type of profile fits the Herschel-based H$_2$ column densities and dust temperature from \citet{Ladjelate2020} and the second type is an analytical profile, described by \citet{Tafalla2002}. We also used two different initial nitrogen abundances.

The chemical model reproduced the observed column densities within its accuracy \citep[one order of magnitude;][]{Sipila2015_model}. However, there are trends in the model results which could be improved: the modelled deuterium fraction of HCO$^+$ is too high, while that of NH$_3$ is too low. The model overproduces the column density of DCO$^+$, and as a result, the deuterium fraction $R_D$(DCO$^+$/HCO$^+$). 

The model overproduces NH$_3$, resulting in a smaller deuterium fraction $R_D$(NH$_2$D/NH$_3$) compared to observations. Previous works also could not produce large values of singly deuterated ammonia in cold cores \citep[e.g.][]{Roueff2005,Harju2017}. The mismatch between the model and the observations may be due to several reasons. First of all, the observations account for the column of the emitting region, which is different for the observed NH$_2$D and NH$_3$ lines due to different excitation conditions, while the model accounts for the column across the whole core. Second, there is an open question of how the deuteration reactions actually proceed; if they proceed via full scrambling (implemented in our work), one expects lower NH$_2$D/NH$_3$ ratios than in the case of  proton hop \citep[see][for the details]{Sipila2019}. Third, in the model we obtain lower $o/p$ ratio both in NH$_3$ and NH$_2$D compared to those adopted to process the observing data: $o/p$(NH$_3$)=0.6--0.8 and $o/p$(NH$_2$D)=1.5--1.8 in the model, while we adopt $o/p$(NH$_3$)=1 and $o/p$(NH$_2$D)=3 \citep[following][]{Harju2017} to obtain the observed total column densities of NH$_3$ and NH$_2$D. The simulated spin ratios are expected to change to statistical ones if proton hop is used \citep{Sipila2019}.

We have the problem of comparing observational and modelling values of CO depletion. We can only estimate the upper limit of CO depletion from observations (Fig. \ref{pic:COfd}). The model shows CO depletion $f_d$=4--7 (see the second row of the plots in Fig.~\ref{pic: model_cn}, \ref{pic: model_emm2}, \ref{pic: model_f}, \ref{pic: model_hmm1}) around the deuteration peak, which is consistent with the observational values. The question of high deuteration along with the low CO depletion is still open.

We compared two density and temperature profiles: profile~1 based on Herschel observations and profile~2 based on analytical radial density distribution. The difference between profile~1 and profile~2 in the models is evident in the Oph-E-MM2, Oph-C-N and Oph-F cores, with profile~2 leading to a significant increase in the column densities of deuterated species. We assume that profile~1, constructed from Herschel observations, does not probe the dense gas on the line of sight where deuteration increases. 
If the physical model would include collapse, the density profile would evolve from the shape of profile~2 to the shape of profile~1, depending on the rate of infall, and would reduce the timescale of chemical evolution \citep[e.g.,][]{Kong2015}.

We compared two different initial nitrogen abundances. Changing the initial nitrogen abundance significantly affected the abundances of nitrogen-bearing species and, as a result, the deuterium fractions. 
The density and temperature profiles better probing the dense gas and based on the observations with higher spatial resolution could also improve the model results for the N-bearing and D-bearing species. Modeling with $X$(N)=7.60$\times$10$^{-5}$ provided the column densities of N$_2$D$^+$, N$_2$H$^+$, and NH$_2$D closer to the observed column densities than $X$(N)=2.14$\times$10$^{-5}$. The assumed static physical conditions limit the model performance.

\section{Conclusions}\label{sec:conclusions}

In this paper we present observational maps towards cold dense cores Oph-C-N, Oph-E-MM2, Oph-H-MM1 and Oph-F and single-pointing observations towards cores Oph-H-MM2 and Oph-I-MM2. We observed the DCO$^+$(2--1), H$^{13}$CO$^+$(2--1), H$^{13}$CO$^+$(1--0), $p$-NH$_2$D(1$_{11}$--1$_{01}$), N$_2$D$^+$(1--0), N$_2$D$^+$(2--1), and N$_2$H$^+$(1--0) lines, measured deuterium fraction in three different species and modelled chemical evolution of the cores. Our findings can be summarised as follows.

\begin{enumerate}

\item We analyse spectra with hyperfine structure and moderate-to-low signal-to-noise ratio. We find the most probable excitation temperature and constrain it to find a realistic estimate of the optical depth. This may result in lower column density estimates compared to an unconstrained fit, but is better than treating the noisy spectra as optically thin. As auxiliary data, we present the parameter maps from our fits: optical depth, central velocity, and velocity dispersion of the lines.
 
\item We presented the column density maps of all six species towards Oph-C-N and Oph-E-MM2, the maps of N$_2$H$^+$, N$_2$D$^+$, NH$_3$, and NH$_2$D towards Oph-F, and the maps of N$_2$H$^+$ and N$_2$D$^+$ towards Oph-H-MM1. The column densities are $\sim10^{12}$~cm$^{-2}$ in N$_2$H$^+$, N$_2$D$^+$, $\sim10^{13}$~cm$^{-2}$ in $p$-NH$_2$D, and $\sim10^{14}$~cm$^{-2}$ in NH$_3$ for all cores. In H$^{13}$CO$^+$ and DCO$^+$, the column densities are $\sim10^{11-12}$~cm$^{-2}$ for Oph-E-MM2 and Oph-C-N.

\item We presented the deuterium fraction maps for three species: $R_D$(N$_2$D$^+$/N$_2$H$^+$) for four cores, $R_D$(NH$_2$D/NH$_3$) for three cores, and $R_D$(DCO$^+$/HCO$^+$) for two cores. While the observed $R_D$(DCO$^+$/HCO$^+$) is similar to deuterium fraction previously observed in other cold cores in C-bearing species (0.014--0.090), the observed $R_D$(N$_2$D$^+$/N$_2$H$^+$) and $R_D$(NH$_2$D/NH$_3$) are very high (0.26--0.81). The high deuterium fraction may be partly due to our line analysis method (in case the excitation temperature of the hydrogenated species was underestimated), but still the studied cores show some of the highest deuterium fractions among cold low-mass cores in the literature. 

\item Deuterium fraction maps $R_D$(DCO$^+$/HCO$^+$), $R_D$(N$_2$D$^+$/N$_2$H$^+$) and $R_D$(NH$_2$D/NH$_3$) show a significant, a factor of 2--10 (median difference is factor of 4.2) difference in the deuterium fraction values for nitrogen-bearing and carbon-bearing species, which confirms the fact that deuterium fraction increases with gas density.

\item We analysed the correlations between the deuterium fractions and the physical parameters of the cores -- H$_2$ column density, gas and dust temperature and the level of turbulence, represented by the non-thermal component of the velocity dispersion. We found no significant correlation with any of the parameters, probably because of the small temperature range covered ($\simeq$2~K) and small area of the $R_D$ maps which do not reach the point of transition to coherence/turbulence.

\item We estimated the upper limit on the CO depletion factor (in all cores) and the lower limit on the ionization degree and cosmic ray ionization rate (in Oph-E-MM2). Both estimates reminded us about a well known paradox of undepleted CO in the dense cores of L1688. The observations of rare CO isotopologues, such as C$^{17}$O or C$^{18}$O both towards the dense cores and undepleted outskirts of the L1688 cloud are needed to measure the CO abundance in the core and study the correlation between deuterium fraction and CO depletion. 

\item We run a model of chemical evolution of the cold cores \citep[pyRate3;][]{Sipila2015_model} trying to reproduce our observation data and constrain the model parameters. We analysed four models with two different gas density profiles and two initial nitrogen elemental abundances, 2.14$\times 10^{-5}$ and 7.60$\times 10^{-5}$, comparing the observed and modelled column densities. We find that the variation in the density profile does not affect the model result significantly. The initial N-abundance, on the contrary, significantly affects the column densities of the N-bearing species and slightly affects the column densities of the C-bearing species. Taking into account the accuracy of the model, in all four variants the modeled column densities agree with the observed ones (within an order of magnitude). However, when it comes to the deuterium fraction, the modelled $R_D$(DCO$^+$/HCO$^+$) is much higher than the observed one; the modelled $R_D$(N$_2$D$^+$/N$_2$H$^+$) is underestimated but close to the observed ones;   the modelled $R_D$(NH$_2$D/NH$_3$) is significantly underestimated compared to the observed ones. The best modelling result for $R_D$(N$_2$D$^+$/N$_2$H$^+$) is obtained with $X$(N$)=2.14\times$10$^{-5}$ and the density profiles based on the Herschel data (profile~1). The best result for $R_D$(DCO$^+$/HCO$^+$) was obtained with $X$(N$)=7.6\times$10$^{-5}$ and the analytical profiles (profile~2). With all four models, NH$_3$ column density is overproduced. 

\item We estimated the chemical ages of the dense cores from the deuterium fraction peaks, the median ages are 2.65$\times10^{5}$, 2.65$\times10^{5}$, 0.82$\times10^{5}$ and 1.80$\times10^{5}$ years for Oph-C-N, Oph-E-MM2, Oph-F and Oph-H-MM1 respectively. The obtained ages are very similar to each other. The chemical ages are not related to dynamical ages, due to the assumed static physical structure.
\end{enumerate}

The spatial distribution of deuterium fraction measured for the three pairs of species in dense cores is presented for the first time and can serve as an observational benchmark for chemical models.

\section*{Acknowledgements}

The authors thank Dr. A.~Pon and Dr. J.~Harju for their help at the earliest stages of the project (preparation of the observing proposals). The authors thank the anonymous referee for the comments that helped to improve the manuscript. This work is based on observations carried out under projects 009-15, and 031-16 with the IRAM~30~m telescope. IRAM is supported by INSU/CNRS (France), MPG (Germany) and IGN (Spain). AP and IP acknowledge the financial support of the Russian Science Foundation project 19-72-00064, AV acknowledges the support of the Ministry of Science and Education of Russia, the FEUZ-2020-0038 project. 

\section*{Data Availability}


The data cubes with the spectra, the spectral line parameter maps, the column density and deuterium fraction maps will be publicly available at the CDS. 



\bibliographystyle{mnras}
\bibliography{example} 




\appendix

\section{Determination of optical depth} \label{depth}
\begin{figure*}
\centering
\includegraphics[scale=0.97]{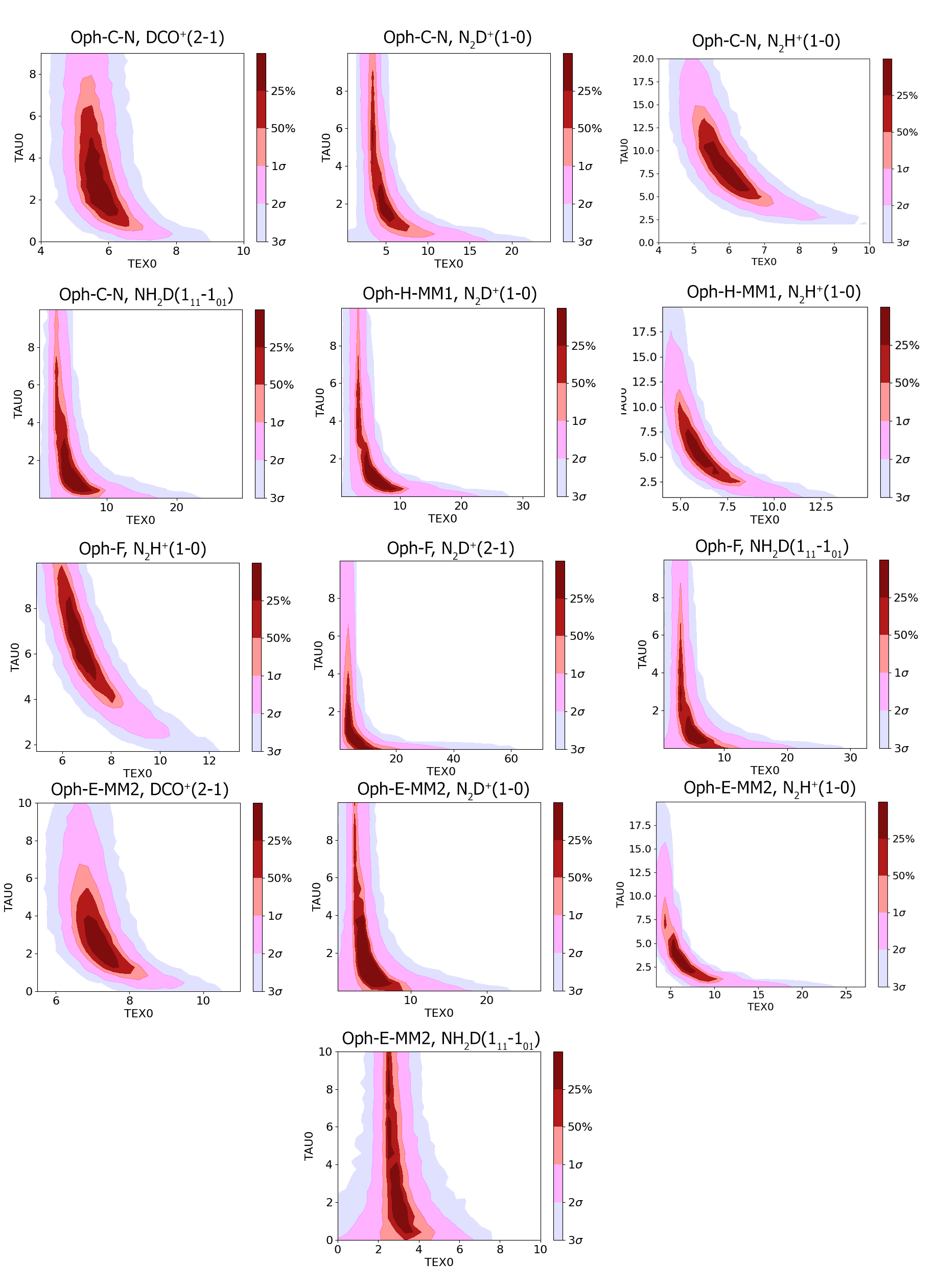}
\caption{Parameter space exploration of the optical depth and the excitation temperature of the observed lines performed with the Monte-Carlo method for Oph-C-N, Oph-E-MM2, Oph-F, and Oph-H-MM1. Abscissa and ordinate show the excitation temperature and the optical depth, respectively. The colour scale shows the uncertainty zones of the normal distribution.}
\label{pic: zone1}
\end{figure*}

\begin{figure*}
\centering
\includegraphics[scale=0.66]{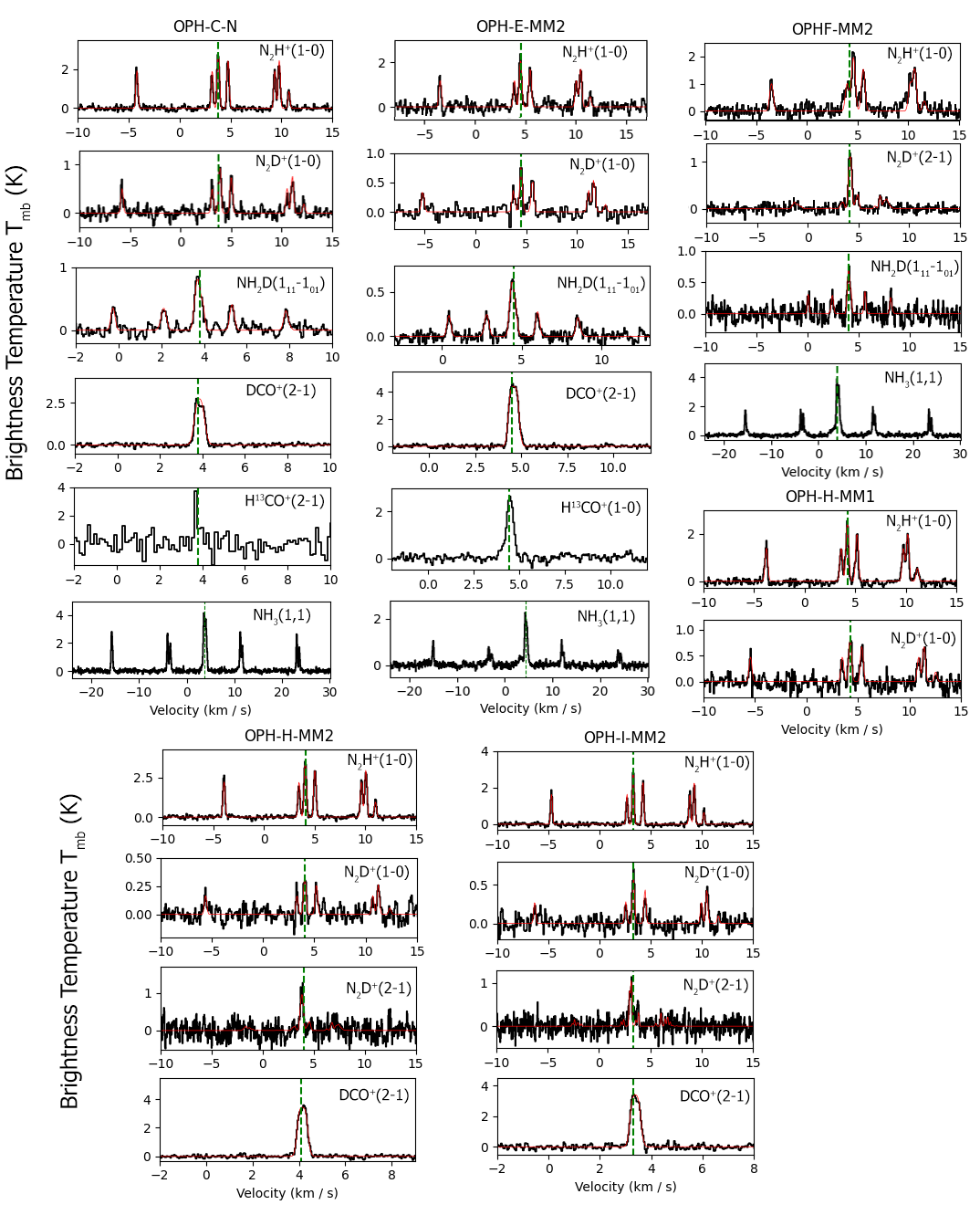}
\caption{The example spectra of the observed lines towards Oph-C-N, Oph-E-MM2, Oph-F, Oph-H-MM1, Oph-I-MM2 and Oph-H-MM2. The black lines show the observed spectrum, the red lines show the synthetic spectra obtained from the hyperfine fitting.}
\label{pic: specCN}
\end{figure*}

Accurate fitting of hyperfine components is essential to measure optical depth. In noisy spectra, it is difficult to accurately determine the optical depth of the line. The optical depth is directly related to the transition excitation temperature, so we decided to fix the transition excitation temperature. To assess the mutual dependence of the optical depth on the excitation temperature, we plotted the probability field of the dependence for the pixels with the best signal-to-noise ratio in the spectral cubes. Figure \ref{pic: zone1} shows these probability fields for each line. This probability field is produced with the Monte-Carlo method, which finds the average value of the parameters and the 3$\sigma$ region, creating a fit of the spectrum many times (we have 10$^5$) with different parameter values. For a fixed value of the excitation temperature, we took the average value found by the Monte-Carlo method. The fixed temperature is found for each line in different cores. With the fixed average value of the excitation temperature, the optical depth for noisy spectra is less than thirty percent, and the fit matches the spectrum, since the difference between the spectrum and the fit is equal to the rms noise level. The spectra with the best signal-to-noise ratios and the fits are shown in Fig.~\ref{pic: specCN}. 


\section{Maps and plots}

\begin{landscape}
\begin{figure*}
\centering
\includegraphics[angle=0,scale=0.48]{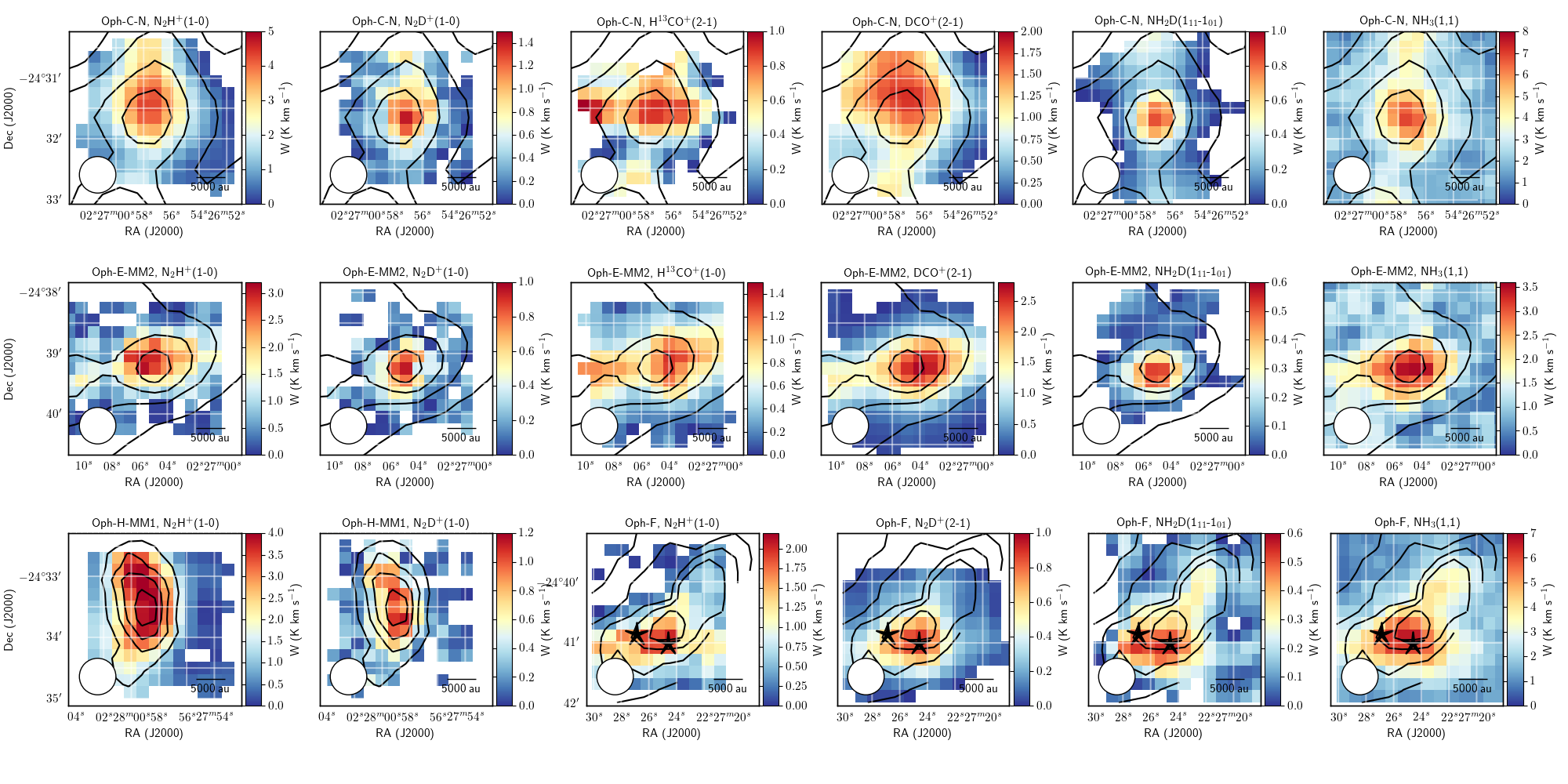}
\caption{Integrated intensities of all studied lines. The contours show the molecular hydrogen column density. The first contour starts at 1.5$\times$10$^{22}$~cm$^{-2}$  with a contour step of 0.5$\times$10$^{22}$~cm$^{-2}$. The beam size is shown in the bottom left corner of each map. The stars show the positions of the YSOs \citep[YLW15, Class~0$+$I, CRBR~2422.8-3423, Class~I;][]{Young1986,Kirk2017,Comeron1993,Bontemps2001} in Oph-F.}
\label{pic: W_map}
\end{figure*}
\end{landscape}

\begin{landscape}
\begin{figure*}
\centering
\includegraphics[angle=0,scale=0.48]{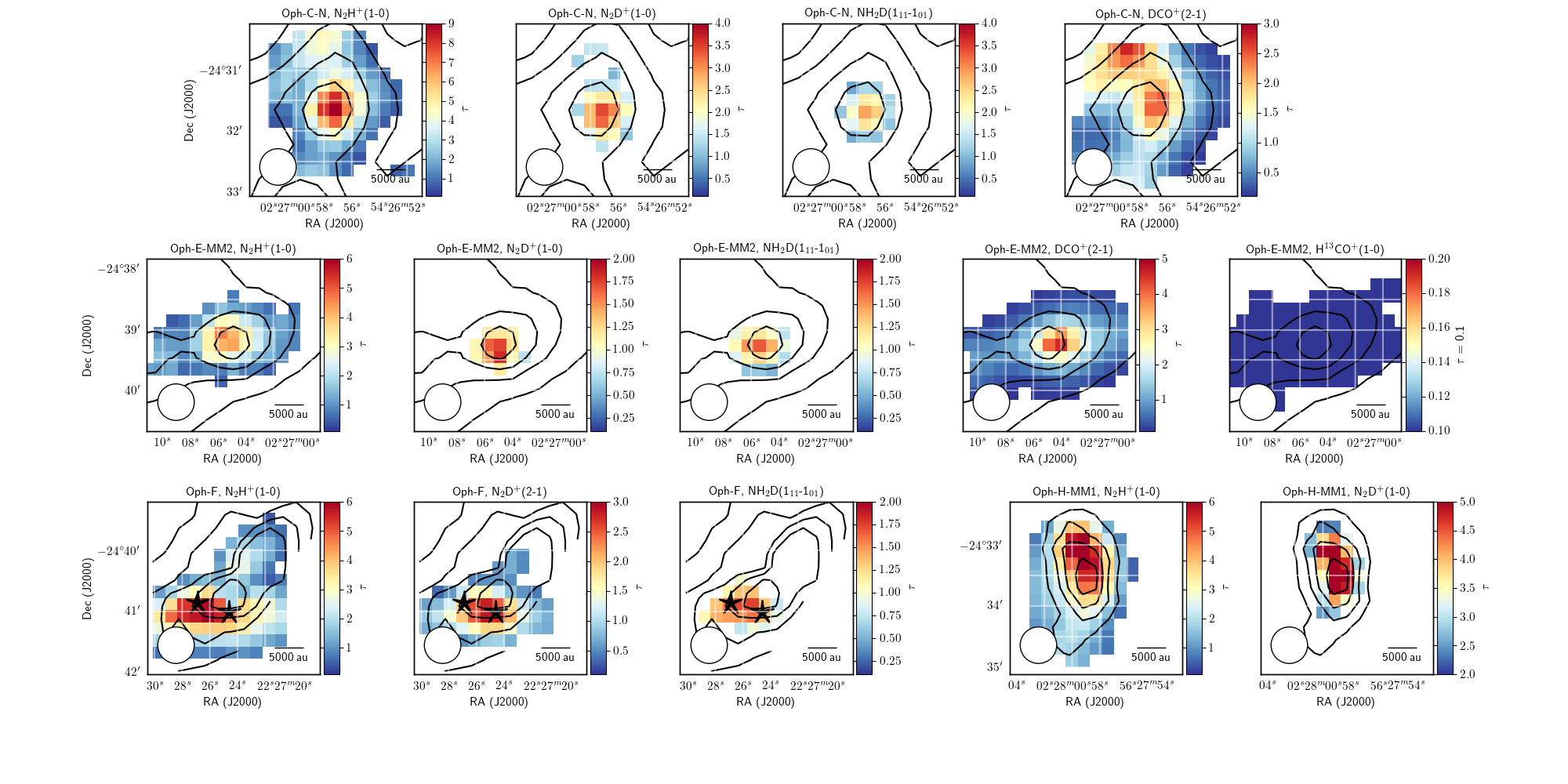}
\caption{Optical depth maps for all studied lines. The contours show the molecular hydrogen column density. The first contour starts at 1.5$\times$10$^{22}$~cm$^{-2}$  with a contour step of 0.5$\times$10$^{22}$~cm$^{-2}$. The beam size is shown in the bottom left corner of each map. The stars show the positions of the YSOs \citep[YLW15, Class~0$+$I, CRBR~2422.8-3423, Class~I;][]{Young1986,Kirk2017,Comeron1993,Bontemps2001} in Oph-F.}
\label{pic: tau}
\end{figure*}
\end{landscape}

\begin{landscape}
\begin{figure*}
\centering
\includegraphics[angle=0,scale=0.48]{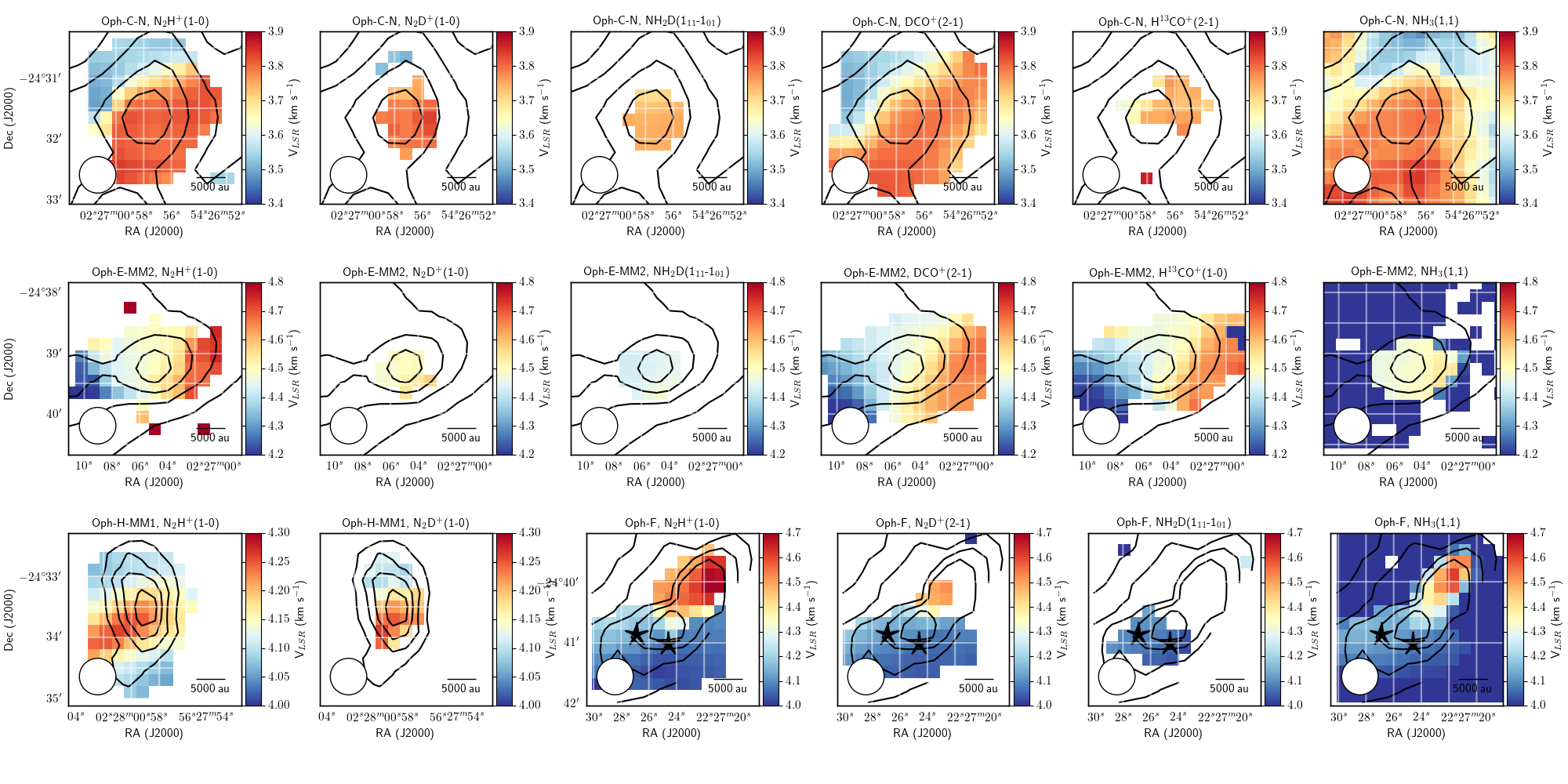}
\caption{Radial velocity maps for all studied lines. The contours show the molecular hydrogen column density. The first contour starts at 1.5$\times$10$^{22}$~cm$^{-2}$  with a contour step of 0.5$\times$10$^{22}$~cm$^{-2}$. The beam size is shown in the bottom left corner of each map. The stars show the positions of the YSOs \citep[YLW15, Class~0$+$I, CRBR~2422.8-3423, Class~I;][]{Young1986,Kirk2017,Comeron1993,Bontemps2001} in Oph-F.} 
\label{pic: vel}
\end{figure*}
\end{landscape}

\begin{landscape}
\begin{figure*}
\centering
\includegraphics[angle=0,scale=0.48]{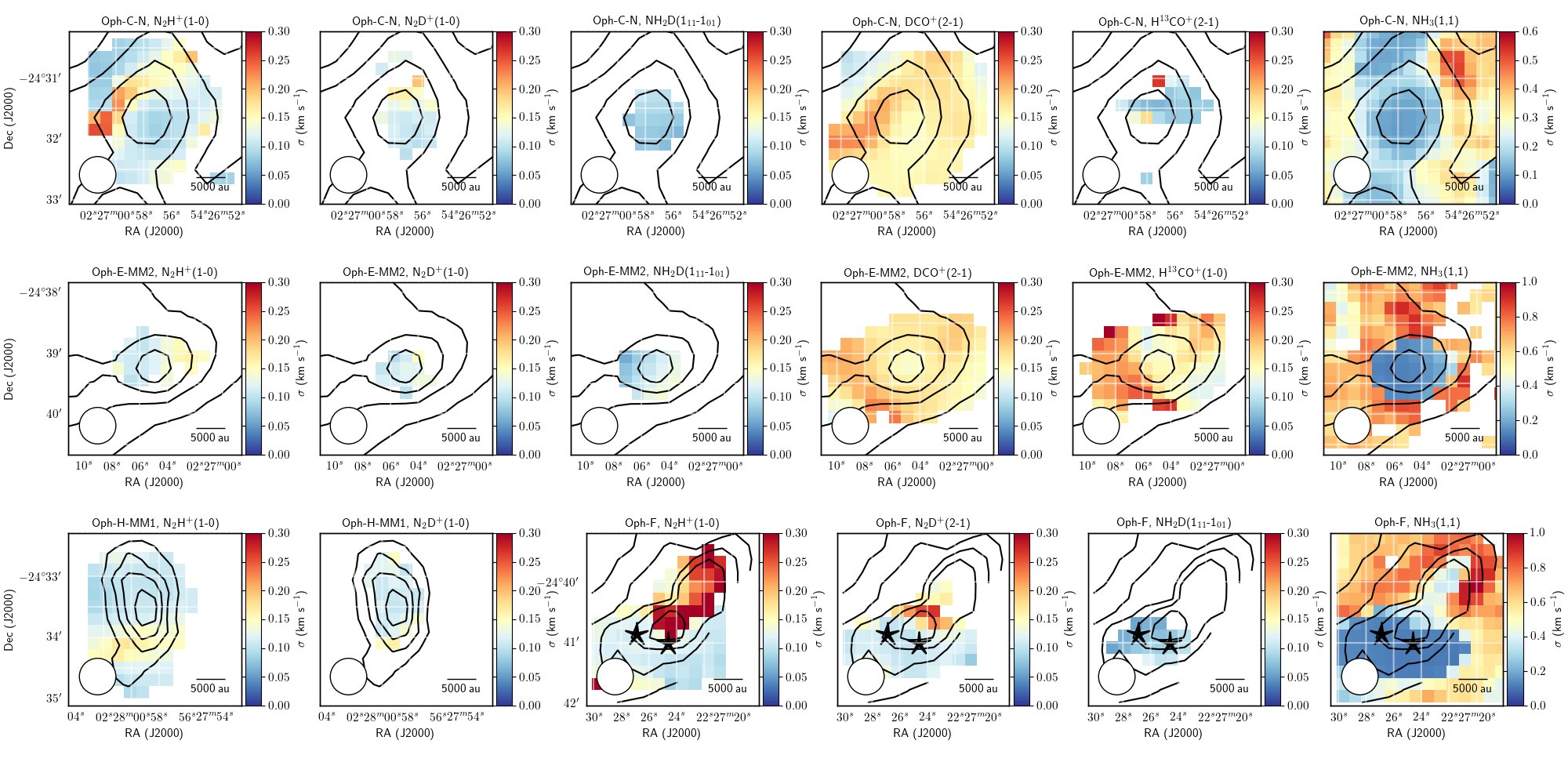}
\caption{Velocity dispersion maps for all studied lines. The contours show the molecular hydrogen column density. The first contour starts at 1.5$\times$10$^{22}$~cm$^{-2}$  with a contour step of 0.5$\times$10$^{22}$~cm$^{-2}$. The beam size is shown in the bottom left corner of each map. The stars show the positions of the YSOs \citep[YLW15, Class~0$+$I, CRBR~2422.8-3423, Class~I;][]{Young1986,Kirk2017,Comeron1993,Bontemps2001} in Oph-F.} %
\label{pic: sigma}
\end{figure*}
\end{landscape}

\begin{figure*}
\centering
\includegraphics[scale=0.4]{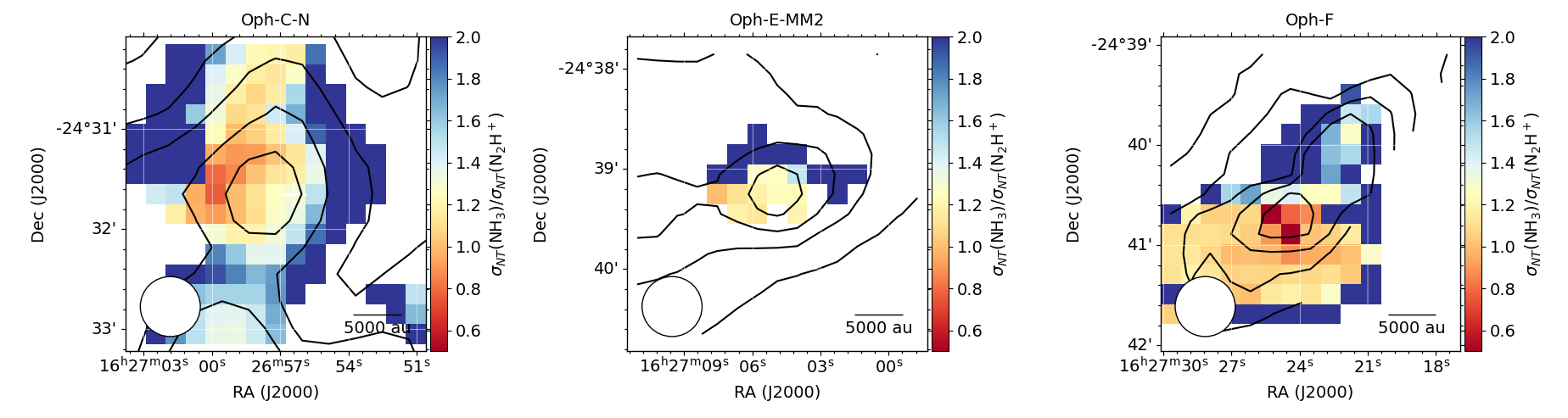}
\caption{NH$_3$(1,1) and N$_2$H$^+$(1--0) non-thermal velocity dispersion components ratio towards Oph-C-N, Oph-E-MM2 and Oph-F cores.}
\label{pic:Sigma_Pineda}
\end{figure*}

\begin{figure*}
\centering
\includegraphics[scale=0.31]{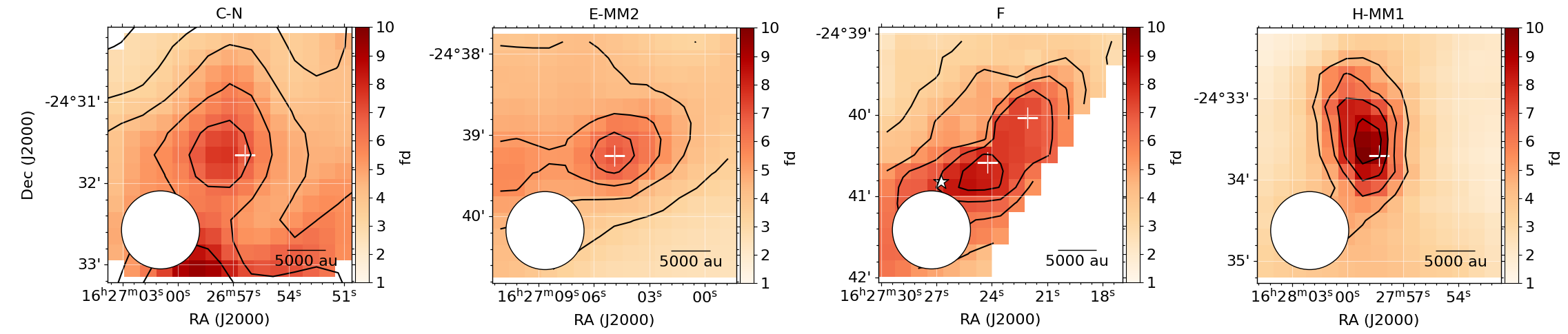}
\caption{The upper limits of the CO depletion factor for the observed cold dense cores. The contours show the molecular hydrogen column density. The first contour starts at 1.5$\times$10$^{22}$~cm$^{-2}$  with a contour step of 0.5$\times$10$^{22}$~cm$^{-2}$. The beam size is shown in the bottom left corner of each map. The white crosses show the positions observed in \citet{Punanova2016}. The stars show the positions of the YSOs \citep[YLW15, Class~0$+$I, CRBR~2422.8-3423, Class~I;][]{Young1986,Kirk2017,Comeron1993,Bontemps2001} in Oph-F.}
\label{pic:COfd}
\end{figure*}

\begin{figure*}
\centering
\includegraphics[scale=0.35]{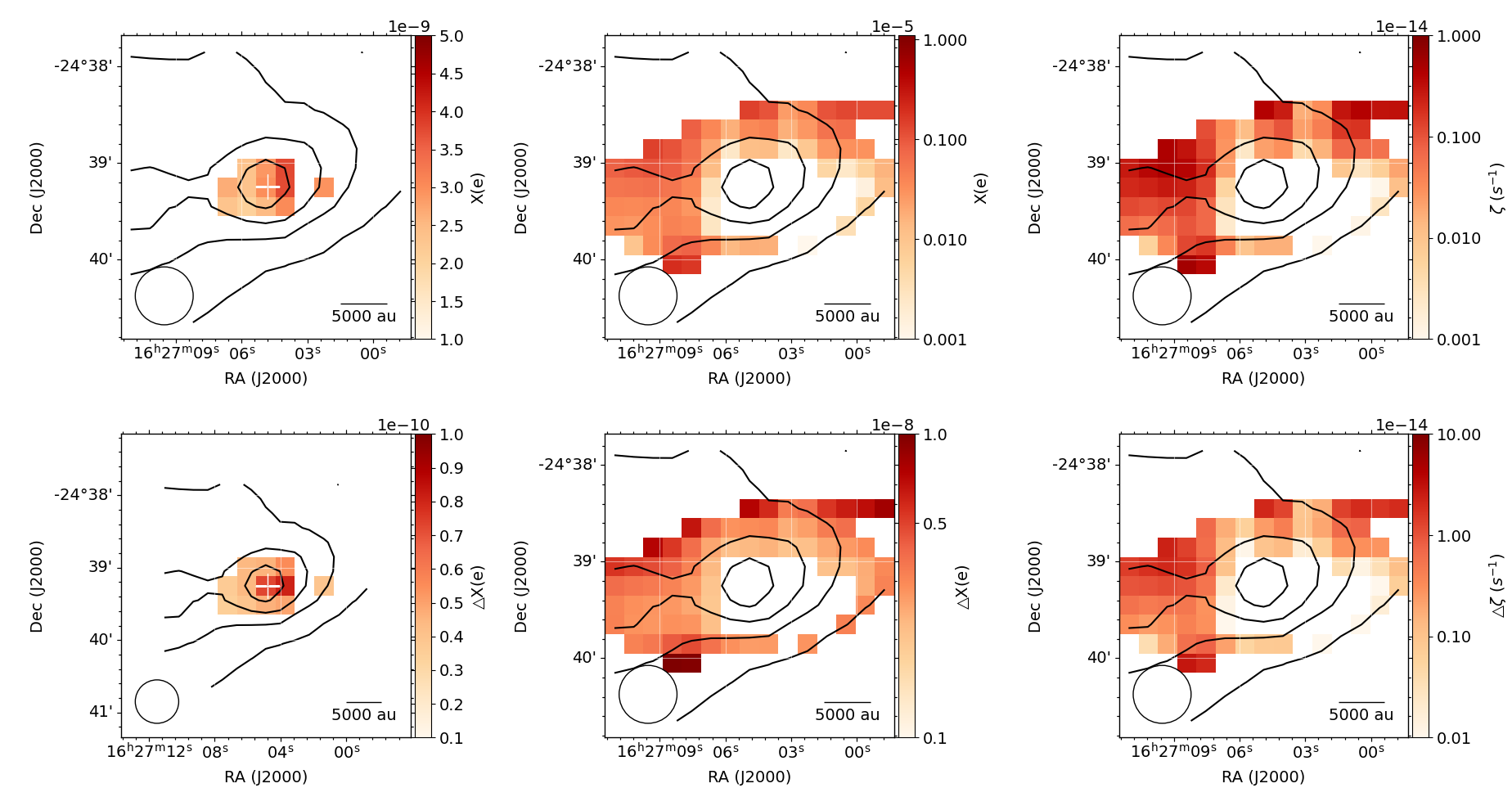}
\caption{Ionization degree $x(e)$ found as abundance of all observed ions (left), ionization degree and ionization rate $\zeta$ based of the observations and analytical model (centre and right). The contours show the molecular hydrogen column density. The first contour starts at 1.5$\times$10$^{22}$~cm$^{-2}$  with a contour step of 0.5$\times$10$^{22}$~cm$^{-2}$. The beam size is shown in the bottom left corner of each map. The white crosses show the positions observed in \citet{Punanova2016}.}
\label{pic:ion}
\end{figure*}

\begin{landscape}
\begin{figure*}
\centering
\includegraphics[angle=270,scale=0.37]{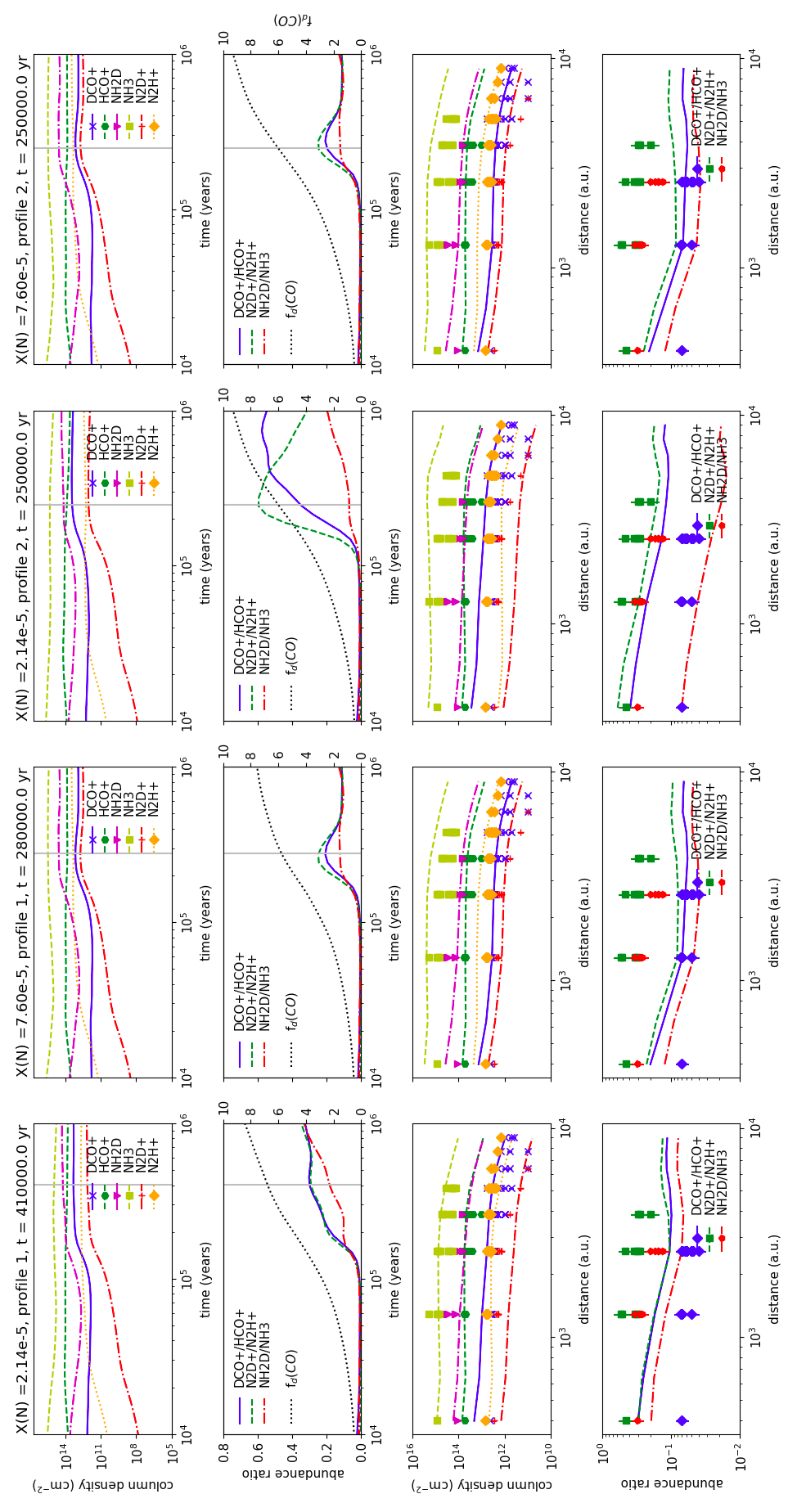}
\caption{The model column densities and deuterium fractions for Oph-C-N. The simulation was performed for two different nitrogen abundances $X$(N)=7.60$\times$10$^{-5}$ and $X$(N)=2.14$\times$10$^{-5}$ and two gas density and dust temperature profiles. Profile~1 is based on observations \citep{Ladjelate2020}, profile~2 is based on the analytical model from \citet{Tafalla2002}. The column densities and deuterium fractions are convolved with the 33.6$^{\prime\prime}$ beam. {\it First (top) row:} temporal evolution of the column densities towards the core centre. {\it Second row:} temporal evolution of the deuterium fractions. {\it Third row:} the column density profiles at the age indicated on top of each column. {\it Fourth (bottom) row:} the deuterium fractions at the age indicated on top of each column. The ages for the profiles are shown by gray vertical lines.} %
\label{pic: model_cn}
\end{figure*} 
\end{landscape}

\begin{landscape}
\begin{figure*}
\centering
\includegraphics[angle=270,scale=0.37]{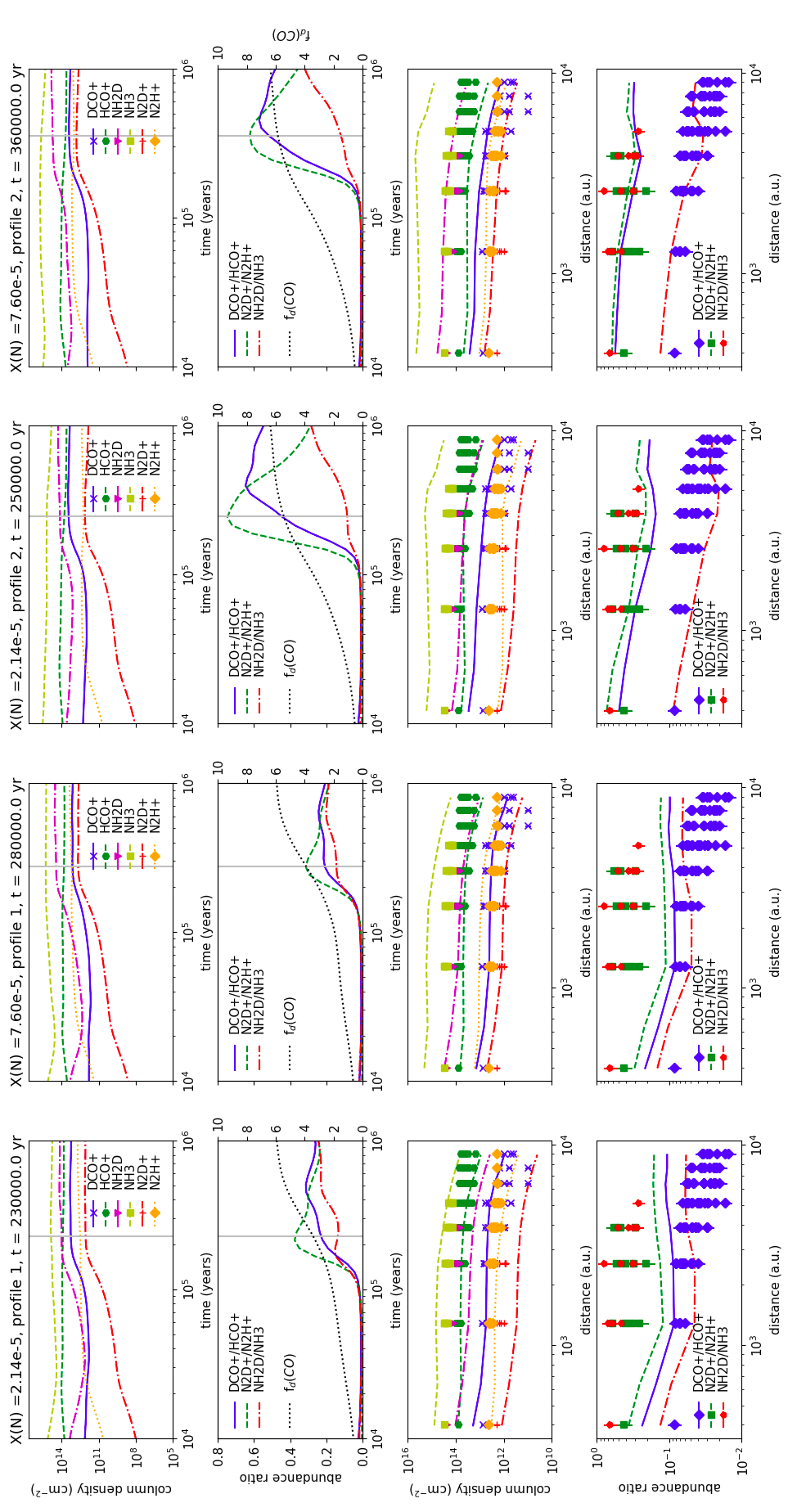}
\caption{The model column densities and deuterium fractions for Oph-E-MM2. The simulation was performed for two different nitrogen abundances $X$(N)=7.60$\times$10$^{-5}$ and $X$(N)=2.14$\times$10$^{-5}$ and two gas density and dust temperature profiles. Profile~1 is based on observations \citep{Ladjelate2020}, profile~2 is based on the analytical model from \citet{Tafalla2002}. The column densities and deuterium fractions are convolved with the 33.6$^{\prime\prime}$ beam. {\it First (top) row:} temporal evolution of the column densities towards the core centre. {\it Second row:} temporal evolution of the deuterium fractions. {\it Third row:} the column density profiles at the age indicated on top of each column. {\it Fourth (bottom) row:} the deuterium fractions at the age indicated on top of each column. The ages for the profiles are shown by gray vertical lines.} %
\label{pic: model_emm2}
\end{figure*}
\end{landscape}

\begin{landscape}
\begin{figure*}
\centering
\includegraphics[angle=270,scale=0.37]{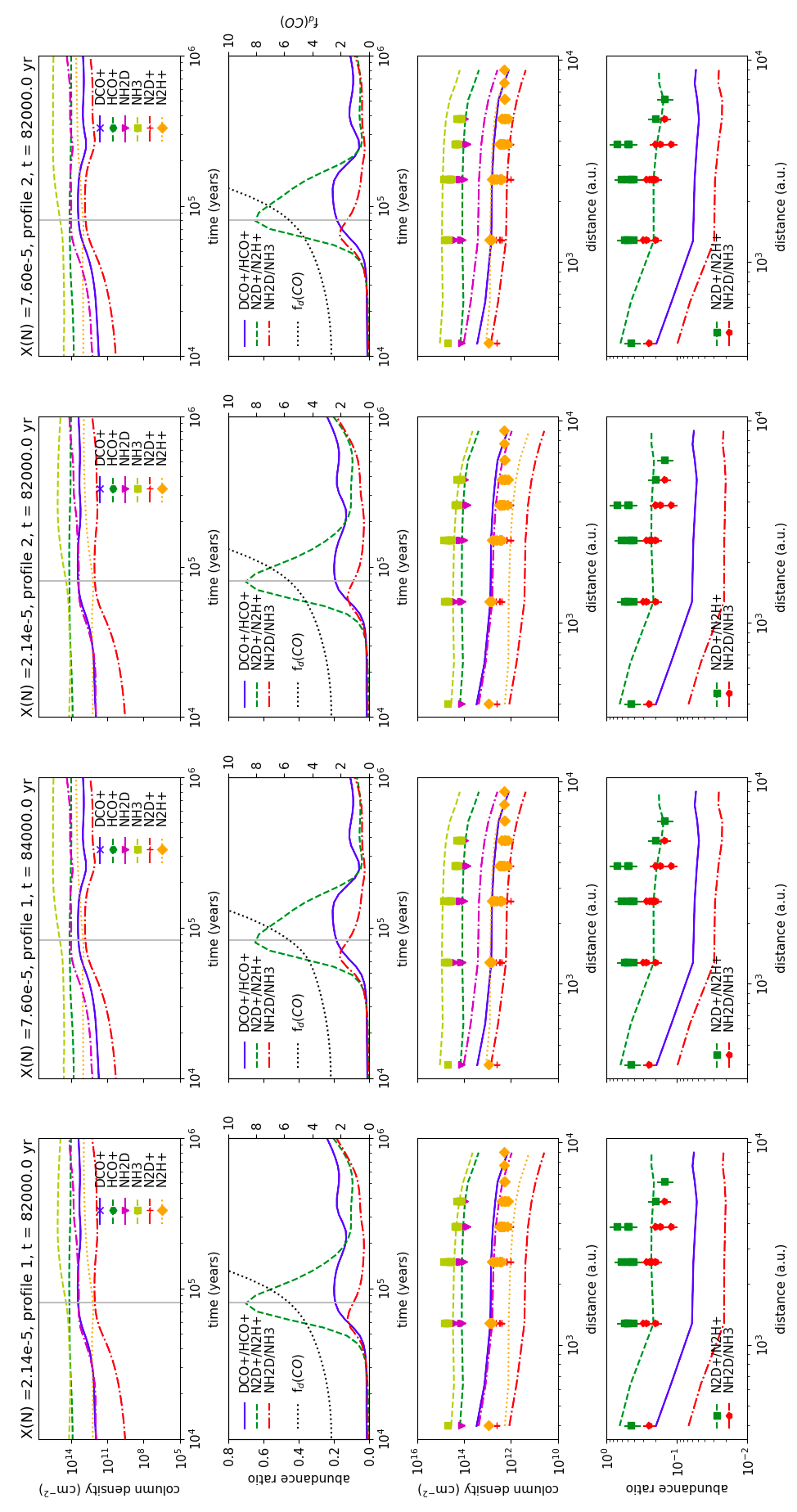}
\caption{The model column densities and deuterium fractions for Oph-F. The simulation was performed for two different nitrogen abundances $X$(N)=7.60$\times$10$^{-5}$ and $X$(N)=2.14$\times$10$^{-5}$ and two gas density and dust temperature profiles. Profile~1 is based on observations \citep{Ladjelate2020}, profile~2 is based on the analytical model from \citet{Tafalla2002}. The column densities and deuterium fractions are convolved with the 33.6$^{\prime\prime}$ beam. {\it First (top) row:} temporal evolution of the column densities towards the core centre. {\it Second row:} temporal evolution of the deuterium fractions. {\it Third row:} the column density profiles at the age indicated on top of each column. {\it Fourth (bottom) row:} the deuterium fractions at the age indicated on top of each column. The ages for the profiles are shown by gray vertical lines.} %
\label{pic: model_f}
\end{figure*}
\end{landscape}

\begin{landscape}
\begin{figure*}
\centering
\includegraphics[angle=270,scale=0.37]{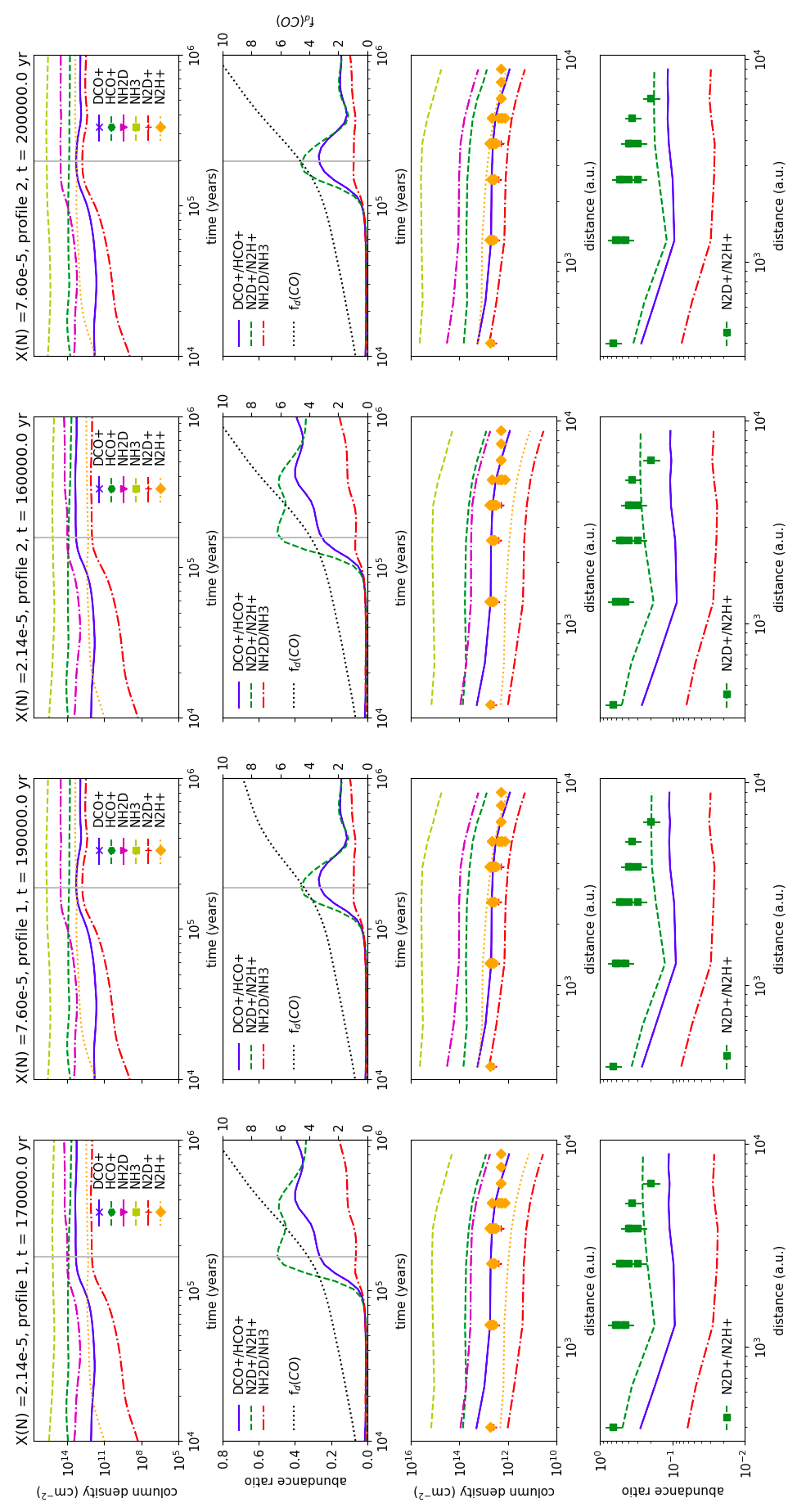}
\caption{The model column densities and deuterium fractions for Oph-H-MM1. The simulation was performed for two different nitrogen abundances $X$(N)=7.60$\times$10$^{-5}$ and $X$(N)=2.14$\times$10$^{-5}$ and two gas density and dust temperature profiles. Profile~1 is based on observations \citep{Ladjelate2020}, profile~2 is based on the analytical model from \citet{Tafalla2002}. The column densities and deuterium fractions are convolved with the 33.6$^{\prime\prime}$ beam. {\it First (top) row:} temporal evolution of the column densities towards the core centre. {\it Second row:} temporal evolution of the deuterium fractions. {\it Third row:} the column density profiles at the age indicated on top of each column. {\it Fourth (bottom) row:} the deuterium fractions at the age indicated on top of each column. The ages for the profiles are shown by gray vertical lines.} %
\label{pic: model_hmm1}
\end{figure*}
\end{landscape}

Here we present the maps of integrated intensity of the lines (Fig.~\ref{pic: W_map}); line parameter maps of $\tau$ (Fig.~\ref{pic: tau}), $V_{\rm LSR}$ (Fig.~\ref{pic: vel}), $\sigma$ (Fig.~\ref{pic: sigma}); the maps of the non-thermal velocity dispersion components ratio $\sigma$(NH$_3$)/$\sigma$(N$_2$H$^+$) (Fig.~\ref{pic:Sigma_Pineda}); the maps of the upper limit of CO depletion (Fig.~\ref{pic:COfd}); ionization degree and ionization rate (Fig.~\ref{pic:ion}). We also present here the plots with the chemical model results (Fig.~\ref{pic: model_cn}, \ref{pic: model_emm2}, \ref{pic: model_f}, \ref{pic: model_hmm1}). 

\onecolumn

\onecolumn

\onecolumn

\onecolumn

\onecolumn

\onecolumn

\onecolumn

\onecolumn

\bsp	
\label{lastpage}
\end{document}